# Designing Practicable Distributed Exchange for Online Communities[*]

Jie Xu[†]    Mihaela van der Schaar[‡]    William Zame[§]

November 1, 2018

**Abstract** In many online systems, individuals provide services for each other; the recipient of the service obtains a benefit but the provider of the service incurs a cost. If benefit exceeds cost, provision of the service increases social welfare and should therefore be encouraged – but the individuals providing the service gain no (immediate) benefit from providing the service and hence have an incentive to withhold service. Hence there is scope for designing a system that improves welfare by encouraging exchange. To operate successfully within the confines of the online environment, such a system should be distributed, practicable, and consistent with individual incentives. This paper proposes and analyzes a simple such system that relies on the exchange of *tokens*; the emphasis is on the design of a protocol (number of tokens and suggested strategies). We provide estimates for the efficiency of such protocols and show that choosing the right protocol will lead to almost full efficiency if agents are sufficiently patient. However, choosing the wrong protocols may lead to an enormous loss of efficiency.

**Keywords** online exchange, token exchange

**JEL Classification D02, D49, E49**

[*]We are grateful to Preston McAfee and to the Associate Editor and referees for helpful comments.
[†]Electrical Engineering, UCLA
[‡]Electrical Engineering, UCLA
[§]Economics,UCLA; corresponding author; zame@econ.ucla.edu

# 1 Introduction

In many online systems, individuals provide services to one another. Napster – now defunct – in which individuals shared music files is surely the most familiar of the systems, but there are many systems and services in current operation. In peer-to-peer networks, individuals share files (Gnutella and Kazaa), or provide computational assistance (Seti@home) or provide content, evaluations and answers to questions (Slashdot, Yahoo Answers). More broadly, many entities (although not necessarily individuals) provide forwarding/routing services to others. In such systems, the recipient of the service obtains a benefit while the provider of the service incurs a cost, which may be quite substantial. Assuming that the benefit exceeds the cost, provision of the service increases social welfare and should therefore be encouraged, but the agents (individuals or entities) providing the service have an incentive to free ride. Empirical studies show that this free-riding problem can be quite severe: in the Gnutella system for instance, almost 70% of users share no files at all (Adar & Hubeman, 2000).

In such settings, there is a great deal of scope for a benevolent designer to improve welfare by careful design of a system that encourages exchange. The extent to which this potential can be realized depends on the environment in which the system operates and on the degree of control the designer can exert. Here we show that a great deal can be achieved by implementing a very simple and practicable system that relies solely on the exchange of intrinsically worthless *tokens* or *fiat money*.[1] We provide estimates for efficiency and an effective procedure for constructing a "good" system. If agents are patient or the benefit/cost ratio is high, efficiency can be made near 1; i.e., trade will take place in almost all interactions.

Our work has a good deal in common with the literature on exchange in the absence of a "double coincidence of wants" and on "money as a medium of exchange", of which Ostroy & Starr (1971) and Kiyotaki & Wright (1989) are the seminal papers (see also especially Green & Zhou (1998) and Zhou (1999)), but differs from that literature in important ways. To understand our work and its relationship to that literature, it is useful to begin by identifying important features of our environment that are common to all the systems mentioned above, and to understand the special restrictions these features impose. The first is that agents interact directly, rather than through a central authority.[2] No central authority means no central recording of transactions and of course no centralized punishments. The second is that the scale is large and interactions are anonymous, so that although agents may interact frequently, they do not interact repeatedly with the same agents, so that reciprocation must be indirect, rather than direct. The third is that, because individuals interact only online, they cannot observe much information about the others with whom they interact; in particular they cannot observe the histories of others. The fourth is that communication in our environment is typically costly (either in direct costs or in terms of delay) or impossible.

---

[1] By "tokens" or "fiat money" we mean the same thing; we use the former term to be consistent with the engineering literature.

The features identified above lead to the first *desideratum* for our system: it should be *completely distributed*, in the sense that the behavior of an agent – in particular, the decision to request/provide service – should depend only on the agents' own history. The second *desideratum* is that the system be *practicable*. At the level of agents, practicability entails minimal exchange of messages. At the level of the designer, practicability entails that the designer initializes the system (in our work this means choosing a *protocol*: providing tokens and recommending strategies), but thereafter takes no actions; in particular, the designer does not monitor the agents or the system (and so cannot limit the number of tokens an agent can hold). Because it seems impossible for the designer to know the precise parameters of the environment (in our setting this means benefit/cost ratios and discount factors), practicability entails that the system should be *robust* to small perturbations of these parameters. Finally, practicability entails *simplicity*. We admit that we cannot provide a working definition of "simplicity" but we hope the results and discussion that follow will make the meaning clearer.

---

[2]Napster and the other systems mentioned above, and many others, *do not* operate through a central authority. Napster, for instance, merely maintained many partial lists (distributed across many servers) of music files available and contact information for subscribers who had these files; users seeking files could simply search these lists and then contact the file-holder directly. For such systems, the absence of a central authority is necessary – a central authority could not handle the amount of traffic generated – and desirable – because a central authority would be vulnerable to attacks that might crash the entire system.

These features and *desiderata* have implications for what is possible in the environments we consider and what distinguishes our work from the literature. If there were a central authority that could record all encounter, then deviations from recommended strategies could be easily detected and punished, but as we have noted above, in our environment there is no central authority or central record-keeping. The fact that agents do not meet in person means that agents cannot inspect the token holdings of others.[3] This makes it impossible to use procedures such as those required in the "two-money theorem" of Kocherlakota (2002), which yields socially optimal outcomes, and in Berensten, Camera & Waller (2007), and other papers, in which the behavior of parties in a match requires that *each party knows the holding of the counter-party*. It also seems impossible to use procedures that require the parties to bargain because bargaining would require either that each party knows the holding (hence the outside option) of the counter-party or that the parties bargain without knowing the holding (hence the outside option) of the counter-party, which (because it would take place in a setting of incomplete information) would require a great deal of communication. As we show in Theorem 5, our desire that the design be robust rules out mixed strategies; similar arguments (discussed in the Conclusion) show that robustness also rules out random or fractional offers. Because packets are essentially indivisible, our environment rules out fractional exchanges, but they would be ruled out by robustness as well. Perhaps more subtly, the impossibility for the designer to monitor the holdings of agents rules out an exogenous upper bound on the number of tokens an agent can hold; contrast Camera & Corbae (1999), Berentsen (2000) and others.[4] As we discuss in Section 3, the absence of an exogenous upper bound on token holdings causes surprising complications (proving that equilibrium exists becomes a difficult task) and surprising implications (the "optimal quantity of money" may be quite different than it is when an exogenous upper bound is imposed).

---

[3]The reader might wonder how agents who do not meet in person can exchange tokens at all, since they can only exchange electronic files, and electronic files would seem to be easily duplicated. In fact, however, there are practicable, secure and private procedures for online token exchange, utilizing hardware or software or both; see Buttyan & Hubaux (2001), Vishnumurthy & Chandrakumar & Sirer (2003) and Ciuffoletti (2010). Similar procedures can also serve as escrow accounts to assure that service that is promised is actually provided and that payment that is promised is actually made.

[4]In a somewhat different vein, Cavalcanti & Wallace (1999A, 1999B) study environments in which *some* agents (banks) can be monitored and can *issue* tokens.

In the environment we consider, a continuum of agents each of whom possesses a unique file that can be duplicated and provided to others. (In the real systems we have in mind, the population is in the tens of thousands or hundreds of thousands, so a continuum model seems a reasonable approximation.) In each period, a fraction of the population is matched; one member of each match (the *client*) must decide whether to request service (provision of a file or forwarding of a packet) and the other (the *server*) must decide whether to provide the service (if requested). The client who receives the service derives a benefit, the server who provides the service incurs a cost. To simplify the analysis we assume here that, except for the uniqueness of the files they possess, all agents are identical, and that all files are equally valuable to receive and equally costly to provide. (We discuss extensions in the Conclusion.) We assume the benefit exceeds the cost, so that social welfare is increased when the service is provided, but that the cost is strictly positive, so that the server has a disincentive to provide it.

We allow the designer to determine a supply of tokens and to recommend strategies (circumstances under which service should be requested or provided); together the token supply and the recommended strategies constitute a *protocol*. We assume that the price of service is fixed at one token; as we have noted above and discuss further in the Conclusion, robustness rules out fractional and random offers. We require that the protocol should induce an equilibrium – the recommended strategies are best replies in the (unique) steady-state distribution – and remain so for small perturbations of the population parameters. There are always degenerate equilibrium protocols in which no service is provided in the steady state. Aside from these degenerate protocols, we show that the requirements of equilibrium and of robustness has strong implications: *all* robust equilibrium protocols are Markov in private token holdings (not history dependent) and *symmetric* (the population plays a pure strategy) and have a particularly simple form: clients request service whenever their token holding is above zero; servers provide service when their token holding is below a certain threshold and do not provide service when the token holding is above that threshold.[5]

---

[5]The fact that the designer recommends the same strategy to all agents is, in part, a consequence of our assumption that agents are identical. If agents were of several types

However, these conclusions do not entirely solve the designer's problem, because he must choose the *right* threshold and the *right* supply of tokens/money. If the threshold is too large, agents will not be willing to continue providing service and accumulating tokens up to the threshold; if the threshold is too small, agents will not be willing to stop providing service and accumulating tokens when they reach the threshold; both of these violate the equilibrium incentive conditions. If the supply of tokens is too small many agents will have no tokens and hence will be unable to purchase service; if the supply of tokens is too large, many agents will be at the threshold and hence will be unwilling to provide service; both of these degrade social welfare. Among all protocols with threshold $K$, there is a unique protocol, in which the supply of tokens is $K/2$, that would be best *if agents always complied with the protocol* – but this protocol need not be an equilibrium. Given the parameters of the population (benefit/cost ratio and discount factor) there is always at least one threshold – and at most two – for which this protocol *is* a robust equilibrium; and we provide estimates for the parameter ranges for which a particular protocol of this form is a robust equilibrium and of the efficiency of such robust equilibrium protocols. A consequence of these estimates is that, as the discount factor tends to 1 or the benefit/cost ratio tends to $\infty$, the efficiency of such equilibrium protocols becomes arbitrarily close to first-best (trade always occurs). However, these protocols are *not* always the optimal equilibrium protocols: it may be more efficient for the designer to choose a higher threshold but to supply fewer tokens. In macro-economic language: $K/2$ *need not be the optimal quantity of money.*

At the risk of redundancy, we would like to emphasize that the environment we study and the questions we ask are different from those in "'standard" monetary theory, and that the differences present new challenges, to which this paper responds. We believe this paper generates a number of insights; among these are:

- the existence of equilibrium protocols is a delicate problem, but all

---

– differing perhaps in the types of files they provide, their costs, the benefits they derive from various types of files, and even the frequency with which they request/provide service – the designer would recommend the same strategy to all agents of a particular type, and equilibrium would again be symmetric.

equilibrium protocols are quite simple;

- the requirement that the system be practicable, and in particular that equilibrium protocols be robust to small perturbations in the population parameters, has significant impact on the nature of equilibrium;

- the most efficient robust equilibrium protocols do not necessarily provide half as many tokens as the selling threshold;

- in order that efficiency close to 1 be achievable, it is necessary that *either* the benefit/cost ratio *or* the discount factor be high *and* that *both* the threshold *and* the supply of tokens be large;

- efficiency can be attained in the limit when the benefit/cost ratio tends to $\infty$ or the discount factor tends to 1

- explicit lower bounds on efficiency can be provided when the benefit/cost ratio is finite and the discount factor is not close to 1;

- there is an effective procedure for constructing 'good' robust equilibrium protocols.

Following a further discussion of the literature below, the remainder of the paper is organized in the following way. Section 2 introduces the model, defines strategies, the steady state value function, best responses, equilibrium and robust equilibrium. Section 3 describes the nature of equilibrium and robust equilibrium. Section 4 discusses existence of robust equilibrium. Section 5 discusses efficiency of equilibrium protocols and shows that asymptotic efficiency can be obtained when agents become infinitely patient or the benefit/cost ratio increases without bound. However, the protocols we identify as asymptotically efficient are not necessarily the most efficient protocols (for fixed population parameters). Section 6 illustrates by a simulation how big the efficiency loss from choosing the wrong protocol can be and proves that efficiency requires a large supply of tokens. Section 7 concludes and offers some directions for further research. Proofs are collected in the Appendix.

**Other Literature**

We have discussed (some of) the literature on "money as a medium of exchange". A somewhat different literature, on "money as memory" (see especially Kocherlakota (1998) and Wallace (2010)) studies the extent to which the role of money can or cannot be served by record-keeping; i.e., when is money *required*? In some sense, the purpose of that literature is nearly the opposite of our purpose here. Record-keeping requires a great deal of centralization; the use of tokens makes possible complete decentralization. An important issue for us is the welfare cost of this complete decentralization, which is discussed in Section 6.

This work also connects to an Electrical Engineering and Computer Science literature that discusses token exchanges in online communities. Some of that literature assumes that agents are compliant, rather than self-interested, and does not treat incentives and equilibrium (Vishnumurthy, Chandrakumar & Sirer 2003), (Buttyan & Hubaux 2003); some of that literature makes use of very different models than the one offered here (Tan & Jarvis 2006) and (Figueiredo, Shapiro & Towsley 2004); and some of the literature is not formal and rigorous, offering simulations rather than theorems (Pai & Mohr 2006). The papers closest to ours are probably Friedman, Halpern and Kash (2006, 2007), which treat somewhat different models. However, these papers seem puzzling in many dimensions and many of the proofs seem mysterious (at least to us).

Another literature to which this work connects is the game-theoretic literature on anonymous interactions. In a context in which interactions were publicly observable, full cooperation (i.e., provision of service) could be achieved at equilibrium by the use of trigger strategies, which deny service in the future to any agent who refuses service in the present. As Kandori (1992) and Ellison (2000) have pointed out, in some contexts, cooperation can be supported even without public observability if agents deny service in the future to *all* agents whenever they have observed an agent who refuses service in the present; in this equilibrium any failure to provide service results in a contagion, producing wider and wider ripples of defection, until no agent provides service. However contagion is not likely to sustain cooperation in the systems of interest to us, because the population is so large (typically comprising tens of thousands or even hundreds of thousands of agents) that an agent is unlikely, in a reasonable time frame, to meet any other agent whose network of past associations overlap with his. (When the population is literally a continuum, no agent *ever* meets any other agent whose network of past associations overlap with his.)

A more relevant literature, of which Kandori (1992) is again the seminal work, uses reputation and social norms as devices as a means of incentivizing cooperation. The work that is closest to ours is Zhang, Park & van der Schaar (2010), which asks which reputation-based systems can be supported in equilibrium and which of these achieve the greatest social efficiency. Because provision of service in their model depends on the reputations of *both* client and server, some central authority must keep track of and verify reputations; hence these systems are not distributed in the sense we use here.

## 2  Model

The population consists of a continuum (mass 1) of infinitely lived agents. Each agent can provide a resource (e.g, a data file, audio file, video file, service) that is of benefit to others but is costly to produce (uploading a file uses bandwidth and time). The benefit of receiving this resource is $b$ and the cost of producing it is $c$; we assume $b > c > 0$.[6] Agents care about current and future benefits/costs and discount future benefits/costs at the constant rate $\beta \in (0, 1)$. Agents are risk neutral so seek to maximize the discounted present value of a stream of benefits and costs.

Time is discrete. In each time period, a fraction $\rho \leq 1/2$ of the population is randomly chosen to be a *client* and matched with a randomly chosen *server*; the fraction $1 - 2\rho$ are unmatched.[7] (No agent is both a client and a server in the same period.) When a client and server are matched, the client chooses whether or not to request service, the server chooses whether or not to provide service (e.g., transfer the file) if requested.

The parameters $b, c, \beta, \rho$ completely describe the environment. Because the units of benefit $b$ and cost $c$ are arbitrary (and tokens have no intrinsic value), only the benefit-cost ratio $r = b/c$ is actually relevant. We consider variations in the benefit-cost ratio $r$ and the discount factor $\beta$, but view the matching rate $\rho$ as immutable.

### 2.1  Tokens and Strategies

In a single interaction between a server and a client, the server has no incentive to provide services to the client. The mechanism we study for creating incentives to provide service involves the exchange of *tokens*. Tokens are indivisible, have no intrinsic value, cannot be counterfeited, and can be stored and transferred without loss. Each agent can hold an arbitrary non-negative finite number of tokens, but cannot hold a negative number of tokens and cannot borrow. We emphasize that our tokens are purely electronic objects and are transferred electronically.

---
[6]If $b \leq c$ there is no social value to providing service; if $c \leq 0$ agents will always be willing to provide service.

[7]We assume that the matching procedure is such that the Law of Large Numbers holds exactly; Duffie & Sun (1997), Als-Ferrer (1999) and Podczek & Puzzello (forthcoming) construct such matching procedures.

The designer creates incentives for the agents to provide or share resources by providing a supply of *tokens* and recommending *strategies* (behavior) for agents when they are clients and servers. At the moment, we allow for strategies that depend on histories but we show that optimal strategies (best responses) depend only on current token holdings.

An *event* describes the particulars of a match at a particular time: whether the agent was chosen to be a client or a server or neither, whether the agent was matched with someone who was willing to serve or to buy, whether the agent received a benefit and surrendered a token or provided service and acquired a token or neither, and the change in the token holding. Write $\epsilon_t$ for an event at time $t$. A *history of length $T$* specifies an initial token holding $m$ and a finite sequence of events $h = (m; \epsilon_0, \epsilon_1, \epsilon_{T-1})$. Write $H_T$ for the set of histories of length $T$, $H = \bigcup_T H_T$ for the set of finite histories. An *infinite history* specifies an initial token holding $m$ and an infinite sequence of events $h = (m; \epsilon_0, \epsilon_1, \ldots)$. We insist that finite/infinite histories be feasible in the sense that net token holdings are never negative (i.e., a request for service by an agent holding 0 tokens will not be honored). Given a finite or infinite history $h$, write $d(h, t)$ for the *change* in token holding at time $t$ and $d^+(h, t), d^-(h, t)$ for the positive and negative parts of $d(h, t)$. Note that $d(h, t) = +1$ if the agent serves, $d(h, t) = -1$ if the agent buys, $d(h, t) = 0$ otherwise. Note also that the token holding at the end of the finite history $h$ is

$$N(h) = m + \sum_{t=0}^{T-1} d(h, t)$$

A *strategy* is a pair $(\sigma, \tau) : H \to \{0, 1\}$; $\tau$ is the *client strategy* and $\sigma$ is the *server strategy*. Following the history $h$, $\tau(h) = 1$ means the client requests service and $\tau(h) = 0$ means the client does not request service; $\sigma(h) = 1$ means the server provides service, $\sigma(h) = 0$ means the server does not provide service. (Note that we require individual agents to follow pure strategies, but we will eventually allow for the possibility that different agents follow different pure strategies, so the population strategy might be mixed.) If service is requested and provided, a single token is transferred from client to server, so the client's holding of tokens decreases by 1 and the server's holding of tokens increases by 1. Tacitly, we assume that a token is transferred if and only if service is provided; like the transfer of tokens itself, this can be accomplished electronically in a completely distributed way.

## 2.2 Steady State Payoffs, Values and Optimal Strategies

Because we consider a continuum population, assume that agents are matched randomly and can observe only their own histories, the relevant state of the system from the point of view of a single agent can be completely summarized by the fraction $\mu$ of agents who do not request service when they are clients and the fraction $\nu$ of agents who do not provide service when they are servers. If the population is in a steady state then $\mu, \nu$ do not change over time. Given $\mu, \nu$, a strategy $(\tau, \sigma)$ determines in the obvious way a probability distribution $P(\tau, \sigma | \mu, \nu)$ over infinite histories $H$. We define the *discounted expected utility* to an agent whose initial token holding is $m$ and who follows the strategy $(\tau, \sigma)$ to be

$$Eu(m, \tau, \sigma | \mu, \nu) = \sum_{h \in H} P(\tau, \sigma | \mu, \nu)(h) \sum_{t=0}^{\infty} \beta^t \left[ d^+(h, t) b - d^-(h, t) c \right]$$

(Here and below, when some of the variables $\beta, b, c, \mu, \nu, \tau, \sigma$ are clearly understood we frequently omit all or some of them; this should not cause confusion.)

Given $\mu, \nu, \tau, \sigma$ and an initial token holding $m$ we define the *value* to be

$$V(m, \mu, \nu, \tau, \sigma) = \sup_{(\tau, \sigma)} Eu(m, \tau, \sigma | \mu, \nu)$$

Discounting implies that the supremum – which is taken over *all* strategy profiles – exists and is at most $b/(1 - \beta)$.

Given $\mu, \nu$ the strategy $(\tau, \sigma)$ is *optimal* or *a best response* for an initial token holding of $m$ if

$$Eu(m, \tau, \sigma | \mu, \nu) \geq Eu(m, \tau', \sigma' | \mu, \nu)$$

for all alternative strategies $\tau', \sigma'$. Because agents discount the future at the constant rate $\beta$, the strategy $(\tau, \sigma)$ is optimal if and only it has the *one-shot deviation property*; that is, there does not exist a finite history $h$ and a profitable deviation $(\tau', \sigma')$ that differs from $(\tau, \sigma)$ following the history $h$ and nowhere else. A familiar and straightforward diagonalization argument establishes that optimal strategies exist and achieve the value; we record this fact below, omitting the proof.

**Proposition 1** *For each $\mu, \nu$ and each initial token holding $m$ there is an optimal strategy $\tau, \sigma$ and*

$$Eu(m, \tau, \sigma | \mu, \nu) = V(m, \mu, \nu, \tau, \sigma)$$

## 2.3 Optimal Strategies

We want to characterize optimal strategies, but before we do, there is a degeneracy that must be addressed. If $\mu = 1$ then no one ever requests service so the choice of whether to provide service is irrelevant; if $\nu = 1$ then no one ever provides service so the choice of whether to request service is irrelevant. In what follows, we sometimes ignore or avoid these degenerate cases, but this should not lead to any confusion.

Fix $\beta, b, c, \mu, \nu$; let $(\tau, \sigma)$ be optimal for the initial token holding $m$. Note that the continuation of $(\tau, \sigma)$ must also be optimal following every history that begins with $m$. If $h$ is such a history and the token holding at $h$ is $n$ then $(\tau, \sigma)$ induces a strategy $(\tau^h, \sigma^h)$ from an initial token holding $n$ that simply transposes what follows $h$ back to time 0, and this strategy must be optimal for the initial token holding of $n$. Conversely, any strategy that is optimal for the initial token holding of $n$ must also be optimal following $h$. It follows that optimal strategies $(\tau, \sigma)$ (whose existence is guaranteed by Proposition 1) *depend only on the current token holding* but are otherwise independent of history; we frequently say such strategies are *Markov* – but note that they are Markov in *individual* token holdings. Write $\Sigma(\mu, \nu, \beta)$ for the set of optimal strategies.

**Theorem 1** *For all $b, c, \beta, \mu, \nu$ with $\nu < 1$, every optimal strategy $(\tau, \sigma)$ has the property that $\tau(n) = 1$ for every $n \geq 1$; i.e. "always request service when possible".*[8]

---

[8] Because a request for service will not be honored when an agent holds 0 tokens, it is irrelevant whether $\tau(0) = 0$ or $\tau(0) = 1$.

In view of Theorem 1, we suppress client strategies $\tau$ entirely, assuming that clients always request service whenever possible. We abuse notation and continue to write $\Sigma(\mu, \nu, \beta)$ for the set of optimal strategies.

We now show that optimal (server) strategies also have a simple form. Say that the (server) strategy $\sigma$ is a *threshold strategy (with threshold $K$)* if

$$\begin{aligned} \sigma(n) &= 1 \quad \text{if} \quad n \leq K \\ \sigma(n) &= 0 \quad \text{if} \quad n > K \end{aligned} \tag{1}$$

We write $\sigma_K$ for the threshold strategy with threshold $K$ and

$$\Sigma = \{\sigma_K : 0 \leq K < \infty\}$$

for the set of threshold strategies.

**Theorem 2** *For each $\mu, \nu, b, c, \beta$ with $\mu < 1$ the set of optimal (server) strategies consists of either a single threshold strategy or two threshold strategies with adjacent thresholds.*

(The assumptions in Theorems 1 and 2 that $\nu < 1$ and $\mu < 1$ avoid the degeneracies previously noted.)

## 2.4 Protocols

The designer chooses a *per capita* supply of tokens $\alpha \in (0, \infty)$ and recommends a strategy to each agent; we allow for the possibility that the designer recommends different strategies to different agents. Because self-interested agents will always play a best response, the designer will recommend only strategies in $\Sigma$; in view of anonymity, it does not matter which agents are recommended to play each strategy, but rather only the fraction of agents recommended to play each strategy. Hence we can identify a recommendation with a *mixed threshold strategy*, which is a probability distribution on $\Sigma$; with the obvious abuse of notation, we view $\gamma$ as a function $\gamma : \mathbb{N}_+ \to [0, 1]$ such that

$$\begin{aligned} \gamma(K) &\geq 0 \text{ for each } K \geq 0 \\ \sum_{K=0}^{\infty} \gamma(K) &= 1 \end{aligned}$$

Write $\Delta(\Sigma)$ for the set of mixed threshold strategies. As usual, we identify the threshold strategy $\sigma_K$ with the mixed strategy that puts mass 1 on $\sigma_K$. Assuming that the designer only recommends best responses (because other recommendations would not be followed), we interpret an element $\gamma \in \Delta(\Sigma)$ as a recommendation that the fraction $\gamma(K)$ play the threshold strategy $\sigma_K$.

A *protocol* is a pair $\Pi = (\alpha, \gamma)$ consisting of a per-capita supply of tokens $\alpha \in (0, \infty)$ and a mixed strategy recommendation $\gamma \in \Delta(\Sigma)$.

## 2.5 Invariant Distributions

If the designer chooses the protocol $\Pi = (\alpha, \gamma)$ and agents follow the recommendation $\gamma$, we can easily describe the evolution of the *token distribution* (the distribution of token holdings). Note that the token distribution must satisfy the two feasibility conditions:

$$\sum_{k=0}^{\infty} \eta(k) = 1 \tag{2}$$

$$\sum_{k=0}^{\infty} k\eta(k) = \alpha \tag{3}$$

Write
$$\mu = \eta(0), \quad \nu = \sum_{\sigma(k)=0} \eta(k)$$

Evidently, $\mu$ is the fraction of agents who have no tokens, hence cannot pay for service, and $\nu$ is the fraction of agents who do not serve (assuming they follow the protocol).

To determine the token distribution next period, it is convenient to think backwards and ask how an agent could come to have $k$ tokens in the next period. There are three possibilities; the agent could have

- $k-1$ tokens in the current period, be chosen as a server, meet a client who can pay for service, and provide service (hence acquire a token);

- $k+1$ tokens in the current period, be chosen as a client, meet a server who provides service, and buy service (hence expend a token);

- $k$ tokens in the current period but and neither provide service nor buy service (hence neither acquire nor expend a token).

Given a recommendation $\gamma$ it is convenient to define $\sigma^\gamma : \mathbb{N}_+ \to [0,1]$ by

$$\sigma^\gamma(n) = \sum_{K=0}^{\infty} \gamma(K)\sigma_K(n)$$

Note that $\sigma^\gamma(n)$ is the fraction of agents in the population who serve when they have $n$ tokens, assuming that agents follow the recommendation $\gamma$ and that the Law of Large Numbers holds exactly in our continuum framework, so $\sigma^\gamma$ is the population strategy. Keeping in mind that token holdings cannot be negative, it is easy to see that the token distribution next period will be

$$\begin{aligned}\eta_+(k) &= \eta(k-1)[\rho(1-\mu)\sigma^\gamma(k-1)] \\ &+ \eta(k+1)[\rho(1-\nu)] \\ &+ \eta(k)[1 - \rho(1-\mu)\sigma^\gamma(k) - \rho(1-\nu)]\end{aligned} \quad (4)$$

where we use the convention $\eta(-1) = 0$.

Given the protocol $\Pi = (\alpha, \gamma)$, the (feasible) token distribution $\eta$ is *invariant* if $\eta_+ = \eta$; that is, $\eta$ is stationary when agents comply with the recommendation $\gamma$. Invariant distributions always exist and are unique.

**Theorem 3** *For each protocol $\Pi = (\alpha, \gamma)$ there is a unique invariant distribution $\eta^\Pi$, which is completely determined by the feasibility conditions (2) and (3) and the recursion relationship*

$$\begin{aligned}\eta^\Pi(k) &= \eta^\Pi(k-1)[\rho(1-\mu)\sigma^\gamma(k-1)] \\ &+ \eta^\Pi(k+1)[\rho(1-\nu)] \\ &+ \eta^\Pi(k)[1 - \rho(1-\mu)\sigma^\gamma(k) - \rho(1-\nu)]\end{aligned} \quad (5)$$

### 2.6 Definition of Equilibrium and Robust Equilibrium

Assuming agents are rational and self-interested, they will comply with a given protocol if and only if compliance is individually optimal; that is, no agent can benefit by deviating from the protocol. To formalize this, fix a protocol $\Pi = (\alpha, \gamma)$, and let $\eta^\Pi$ be the unique invariant distribution. Write

$$\mu^\Pi = \eta^\Pi(0) \ , \ \nu^\Pi = \sum_{\sigma(k)=0} \eta^\Pi(k)$$

for the fraction of agents who have no tokens and the fraction of agents who do not serve (in the invariant distribution induced by $\Pi$), respectively. We say $\Pi = (\alpha, \gamma)$ is an *equilibrium protocol* if $\sigma_K$ is an optimal strategy (given given $\mu^\Pi, \eta^\Pi$) whenever $\gamma(K) > 0$. That is, $\gamma$ puts positive weight only on threshold strategies that are optimal, given the invariant distribution that $\Pi$ itself induces.

Using the one step deviation principle, we can provide a useful alternative description of equilibrium in terms of the value function $V$. As noted before, because optimal strategies exist and are Markov, we may unambiguously write $V_k$ for the value following *any* history at which the agent has $k$ tokens. (The value function depends on the population data $\mu, \nu$ and on the environmental parameters $b, c, \beta$; but there should be no confusion in suppressing those here.)

Fix any Markov strategy $\sigma$. In order for $\sigma$ to be optimal, it is necessary and sufficient that it achieves the value $V_\ell$ following *every* token holding $\ell$. Expressed in terms of *current* token holdings and *future* values, and taking into account how behavior in a given period affects the token holding in the next period, this means that $\sigma$ is optimal if and only if it satisfies the following system of equations:

$$\begin{aligned}
V_0 &= \rho\sigma(0)[(1-\mu)(-c+\beta V_1) + \mu\beta V_0] \\
&\quad + \rho[1-\sigma(0)]\beta V_0 + (1-2\rho)\beta V_0
\end{aligned}$$

$$\begin{aligned}
V_k &= \rho[(1-\nu)(b+\beta V_{k-1}) + \nu\beta V_k] \\
&\quad + \rho\sigma(k)[(1-\mu)(-c+\beta V_{k+1}) + \mu\beta V_k] \\
&\quad + \rho[1-\sigma(k)]\beta V_k + (1-2\rho)\beta V_k \\
&\text{for each } k > 0
\end{aligned} \tag{6}$$

Applying this observation to the threshold strategy $\sigma_K$ and carrying out the requisite algebra, we conclude that $\sigma_K$ is optimal if and only if

$$-c + \beta V_{k+1} \geq \beta V_k \text{ if } k \leq K \qquad (7)$$
$$-c + \beta V_{k+1} \leq \beta V_k \text{ if } k > K \qquad (8)$$

(If it seems strange that $\alpha, \gamma$ do not appear in these inequalities, remember that the value depends on the invariant distribution $\eta^\Pi$, which in turn depends on $\alpha$ and on $\gamma$.)

Given a benefit/cost ratio $r > 1$ and a discount factor $\beta < 1$, write $EQ(r, \beta)$ for the set of protocols $\Pi$ that constitute an equilibrium when the benefit/cost ratio is $r$ and the discount factor is $\beta$. Conversely, given a protocol $\Pi$ write $E(\Pi)$ for the set $\{(r, \beta)\}$ of pairs of benefit/cost ratios $r$ and discount factors $\beta$ such that $\Pi$ is an equilibrium protocol when the benefit/cost ratio is $r$ and discount factor is $\beta$. Note that $EQ, E$ are correspondences (which might have empty values) and are inverse to each other.

Given $r, \beta$ we say that $\Pi$ is a *robust equilibrium* if $(r, \beta)$ belongs to the interior of $E(\Pi)$; i.e., there is some $\varepsilon > 0$ such that $\Pi \in EQ(r', \beta')$ whenever $|r'-r| < \varepsilon$ and $|\beta'-\beta| < \varepsilon$. Write $EQR(r, \beta)$ for the set of robust equilibrium protocols for the benefit/cost ratio $r$ and discount factor $\beta$ and $ER(\Pi)$ for the set $\{(r, \beta)\}$ of pairs of benefit/cost ratios for which $\Pi$ is a robust equilibrium. Note that $EQR, ER$ are correspondences (which might have empty values) and are inverse to each other.

## 3   Equilibrium and Robust Equilibrium

We first describe the nature of equilibrium and robust equilibrium and then use that description to show that robust equilibria exist. The crucial fact about equilibrium is that the strategy part of an equilibrium protocol can involve mixing over at most two (thresholds and that these thresholds must be adjacent; the crucial fact about robust equilibrium is that the strategy cannot involve strict mixing at all but must rather be a pure strategy.

**Theorem 4** *For each benefit/cost ratio $r > 1$ and discount factor $\beta < 1$ the set $EQ(r, \beta)$ is either empty or consists of protocols that involve only (possibly degenerate) mixtures of two threshold strategies with adjacent thresholds.*

**Theorem 5** *If $\Pi = (\alpha, \sigma)$ is a robust equilibrium then $\sigma$ is a pure threshold strategy.*

The existence of equilibrium or robust equilibrium does not seem at all obvious (and our proof is not simple). For both intuition and technical convenience, it is convenient to work "backwards": rather than beginning with population parameters $r, \beta$ and looking for protocols $\Pi$ that constitute an equilibrium for those parameters, we begin with a protocol $\Pi$ and look for population parameters $r, \beta$ for which $\Pi$ constitutes an equilibrium. That is, we do not study the correspondences $EQ(r, \beta)$ and $EQR(r, \beta)$ directly, but rather the inverse correspondences $E(\Pi)$ and $ER(\Pi)$. (This is easier for several reasons, one of which is that the latter correspondences are *always* non-empty.)

To give an intuitive understanding of the difficulty and how we overcome it, fix a protocol $\Pi = (\alpha, \sigma)$ and let $\eta^\Pi$ be the invariant distribution. Because we will eventually want to find a robust equilibrium, we assume $\sigma$ is a threshold strategy: $\sigma = \sigma_K$. To look for population parameters $r, \beta$ for which $\Pi$ is an equilibrium, let us fix $r$ and let $\beta$ vary. (Alternatively, we could fix $\beta$ and let $r$ vary, or vary both $\beta, r$ simultaneously, but the intuition is most easily conveyed by fixing $r$ and letting $\beta$ vary.) As we have already noted, the invariant distribution $\eta^\Pi$, and hence $\mu^\Pi, \nu^\Pi$, depends only on $\Pi$. If $\beta$ is close to 0, an agent will have little incentive to acquire tokens, but the incentive to acquire tokens increases as $\beta \to 1$. It can be shown that there is a smallest discount factor $\beta_L(\Pi)$ with the property that an agent whose discount factor is at least $\beta_L(\Pi)$ will be willing to continue providing service until he has acquired $K$ tokens. This is not enough, because $\sigma_K$ will only be incentive compatible if the agent is also willing to *stop* providing service *after* he has acquired $K$ tokens. However, it can also be shown that there is a largest discount factor $\beta_H(\Pi)$ for which the agent is willing to stop providing service after he has acquired $K$ tokens, and that $\beta_L(\Pi) < \beta_H(\Pi)$. (Recall that $r, \Pi$ are fixed.) For every discount factor $\beta$ in the closed interval $[\beta_L(\Pi), \beta_H(\Pi)]$, the protocol $\Pi$ is an equilibrium when the population parameters are $r, \beta$; that is, $(r, \beta) \in E(\Pi)$. From this it can be shown that for every discount factor $\beta$ in the interval $(\beta_L(\Pi), \beta_H(\Pi))$, the protocol $\Pi$ is a robust equilibrium when the population parameters are $r, \beta$; that is, $(r, \beta) \in ER(\Pi)$. Similarly, we can hold $\beta$ fixed and let $r$ vary from 1 to $\infty$, construct the corresponding intervals $[r_L(\Pi), r_H(\Pi)]$ with $r_L(\Pi) < r_H(\Pi)$ and then show that for every benefit/cost ratio $r$ in the open interval $(r_L(\Pi), r_H(\Pi))$ the protocol $\Pi$ is a robust equilibrium when the population parameters are $r, \beta$; that is, $(r, \beta) \in ER(\Pi)$. This is the content of Theorem 6 below.

Applying this procedure for every protocol yields a family $\{ER(\Pi)\}$ of non-empty open sets of parameters $r, \beta$ for which robust equilibria exist. However our work is not done; we would also like to know that $\{ER(\Pi)\}$ *covers* a big enough set of population parameters. For instance, we want to know that for each $r > 1$ there is a $\beta^* < 1$ such that $\{ER(\Pi)\}$ covers the set $\Pi(r, \beta) : \beta > \beta^*\}$; this would mean that for each $r > 1$ and $\beta > \beta^*$ there is a protocol $\Pi$ that constitutes a robust equilibrium for these population parameters. Proving that this is so is not at all easy, and requires deriving some special properties of protocols of the form $\Pi_K = (K/2, \sigma_K)$; this is the content of Theorem 7 below.

It is natural to ask why our proof seems (and is) so much more complicated than existence proofs in the literature, such as in Berentsen (2000). The answer is that the literature assumes that there is an *exogenous upper limit* $K^*$ on the number of tokens any agent can hold. As discussed above, this assumption makes it relatively easy to show that equilibrium exists: Fix the benefit/cost ratio fix $r > 1$ and an arbitrary $\alpha > 0$ and consider the protocol $(\alpha, \sigma_{K^*})$. As above, an agent whose discount factor $\beta$ is at least $\beta_L(\alpha, \sigma_{K^*})$ will provide service until he has acquired $K^*$ tokens, and will *stop* providing service *after* he has acquired $K^*$ tokens because, by assumption, he *cannot hold more than $K^*$ tokens*. Hence $(r, \beta) \in E(\alpha, \sigma_{K^*})$ for *every* $\beta \geq \beta_L(\Pi)$ and $(r, \beta) \in ER(\alpha, \sigma_{K^*})$ for *every* $\beta > \beta_L(\alpha, \sigma_{K^*})$. As we have noted in the Introduction, assuming an exogenous upper bound on token holdings does not seem realistic in the environments we consider. However, even if a large exogenous upper bound $K^*$ *were* imposed, our approach would yield much more, because we can *construct* robust equilibria for discount factors much smaller than $\beta_L(\alpha, \sigma_{K^*})$.

**Theorem 6** *Fix a protocol $\Pi = (\alpha, \sigma_K)$.*

(i) *For each benefit/cost ratio $r > 1$, the set $\{\beta : \Pi \in EQ(r, \beta)\}$ is a non-degenerate closed interval $[\beta_L(\Pi), \beta_H(\Pi)]$ whose endpoints are continuous functions of $r$.*

(ii) *For each discount factor $\beta < 1$, the set $\{r : \Pi \in EQ(r, \beta)\}$ is a non-degenerate closed interval $[r_l(\Pi), r_H(\Pi)]$ whose endpoints are continuous functions of $\beta$.*

These results are illustrated for $\alpha = 1/4$ in in Figures 1 and 2. (Figure 1 may give the impression that the intervals for successive values of $K$ do not overlap, but as Figure 2 illustrates, they actually *do* overlap; the overlap is masked by the granularity of the Figure. However, as we have already said, we do not assert that overlapping of intervals for successive values of $K$ is a general property.)

For the special protocols $\Pi_K = (K/2, \sigma_K)$, in which the supply of tokens is exactly half the selling threshold, we can say a great deal more. In particular, the intervals corresponding to successive values of the threshold overlap but are not nested. As suggested above, this is exactly what we need to guarantee that equilibria exist (for sufficiently large values of the parameters $r, \beta$).

**Theorem 7**

(i) *For each fixed benefit-cost ratio $r > 1$, successive $\beta$-intervals overlap but are not nested:*

$$\beta_L(\Pi_{K-1}) < \beta_L(\Pi_K) < \beta_H(\Pi_{K-1}) < \beta_H(\Pi_K)$$

*for every threshold $K$. Moreover*

$$\lim_{K \to \infty} \beta_L(\Pi_K) = 1$$

*In particular, there is some $\beta^* < 1$ such that $EQR(r, \beta) \neq \emptyset$ for all $\beta > \beta^*$.*

(ii) *For each fixed discount factor $\beta < 1$, successive $r$-intervals overlap but are not nested:*

$$r_L(\Pi_{K-1}) < r_L(\Pi_K) < r_H(\Pi_{K-1}) < r_H(\Pi_K)$$

*for every threshold $K$. Moreover*

$$\lim_{K \to \infty} r_L(\Pi_K) = \infty$$

*In particular, there is some $r^* > 1$ such that $EQR(r, \beta) \neq \emptyset$ for all $r > r^*$.*

It follows from Theorem 7 that, as $K \to \infty$, the left-hand end-points $\beta_L(\Pi_K) \to 1$, so *a fortiori* the lengths of $\beta$-intervals shrink to 0. It is natural to guess that the lengths of these intervals shrink *monotonically* to 0, and simulations suggest that this guess is correct, but we have neither a proof nor a good intuition that this is actually true. We also guess that the lengths of $r$-intervals shrink monotonically, but again we have neither a proof nor a good intuition that this is actually true.

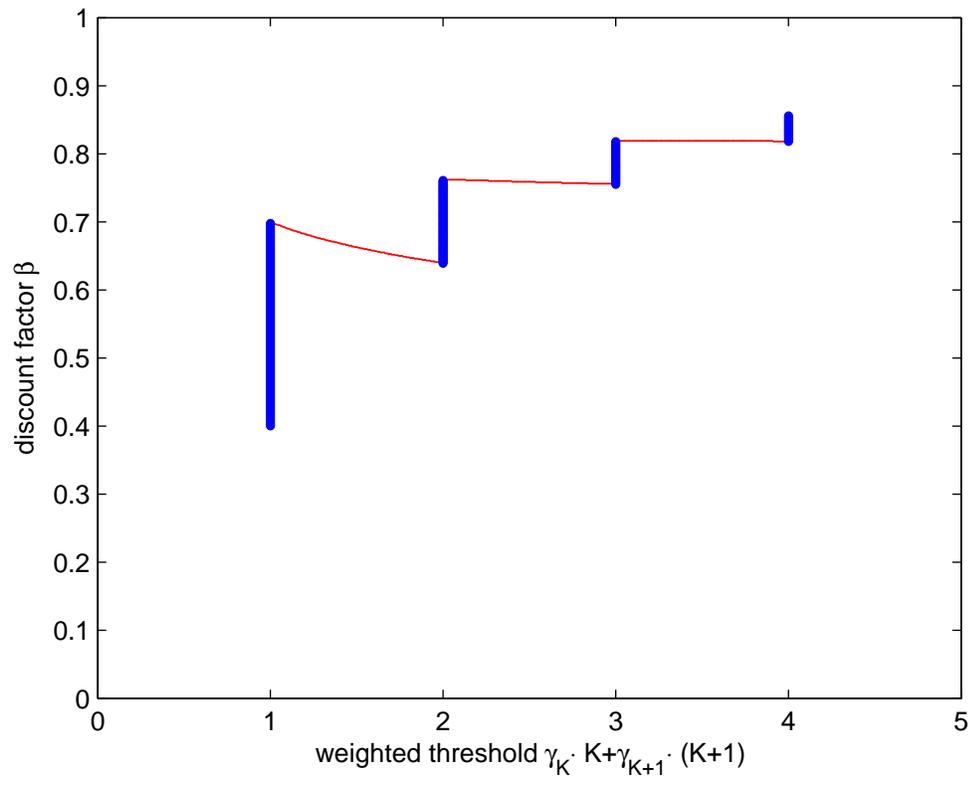

Figure 1: Pure and Mixed Equilibrium: $\alpha = 1/4$
(blue (thick) - pure equlibrium; red (thin) - mixed equilibrium)

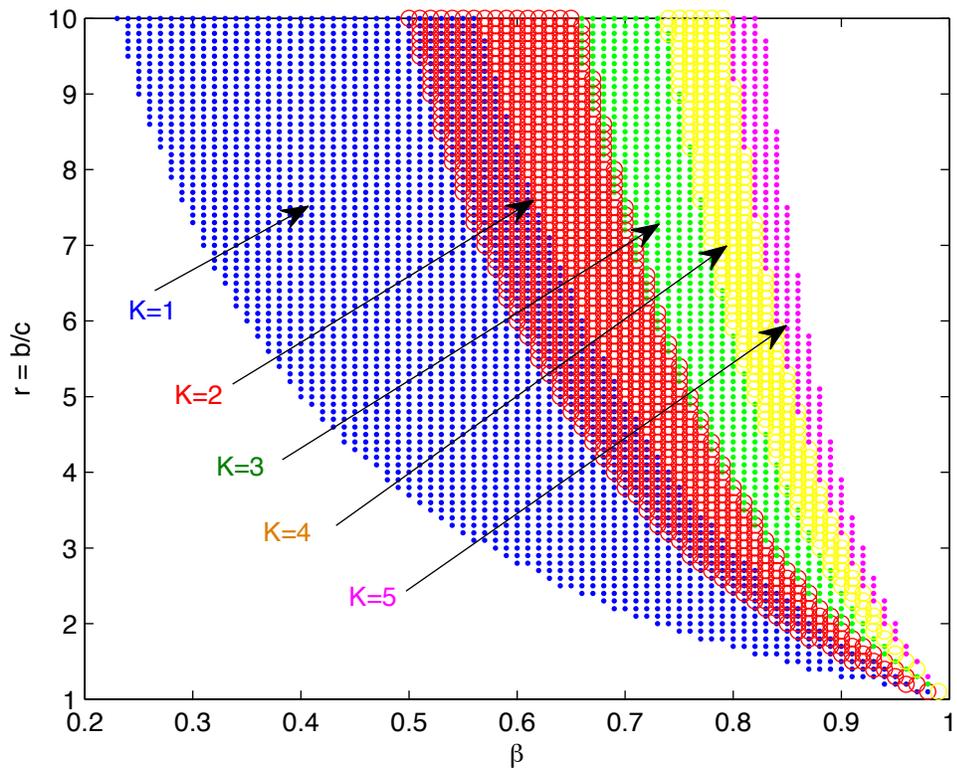

Figure 2: Threshold Equilibrium: $\alpha = 1/4$

## 4 Efficiency

If agents were compliant (rather than self-interested), the designer could simply instruct them to provide service at every meeting and they would comply, so the per capita social gain in each period would be $\rho(b-c)$. If agents follow the protocol $\Pi = (\alpha, \sigma_K)$ then service will be provided only in those meetings where the client can buy service and the server is willing to provide service, so the per capita social gain in each period will be $\rho(b-c)(1-\mu^\Pi)(1-\nu^\Pi)$. Hence we define the *efficiency* of the protocol $\Pi$ to be
$$\text{Eff}(\Pi) = (1-\mu^\Pi)(1-\nu^\Pi)$$
In general it seems hard to determine the efficiency of a given protocol or to compare the efficiency of different protocols. However, we can provide efficiency bounds for protocols that utilize a given threshold strategy $\sigma_K$ and compute the precise efficiency of the protocols $\Pi_K$.[9]

**Theorem 8** *For each $\alpha \in (0, \infty)$, each threshold $K$ and all values of the population parameters we have:*

*(i)* $\text{Eff}(\alpha, \sigma_K) \leq 1 - \frac{1}{2\lceil \alpha \rceil + 1}$

*(ii)* $\text{Eff}(\alpha, \sigma_K) \leq \text{Eff}(\Pi_K)$

*(iii)* $\text{Eff}(\Pi_K) = \left(1 - \frac{1}{K+1}\right)^2 = \left(\frac{K}{K+1}\right)^2$

Two implications of Theorem 8 are immediate. The first is that, in order that a (threshold) protocol achieve efficiency near 1 it is necessary that it provide a large number of tokens *and also* that it prescribe a high selling threshold. Put differently: to yield full efficiency in the limit it is *not* enough to increase the number of tokens without bound or to increase the threshold without bound – *both* must be increased without bound. The second is that the protocols $\Pi_K$ that provide $K/2$ tokens per capita are the most efficient protocols that utilize a given threshold strategy $\sigma_K$.

We caution the reader, however, that the protocols $\Pi_K$ *need not be* equilibrium *protocols*, and it is (robust) equilibrium protocols that we seek.

---

[9] Berentsen (2000) derives similar results in a different model, with Poisson arrival rates.

However, it follows immediately from Theorem 7 that whenever agents are sufficiently patient or the benefit-cost ratio is sufficiently large (or both), then *some* protocol $\Pi_K$ is an equilibrium for large $K$, and hence that nearly efficient equilibrium protocols always exist.

**Theorem 9**

(i) for each fixed discount factor $\beta < 1$

$$\liminf_{r \to \infty} \sup\{\text{Eff}(\Pi_K) : \Pi_K \in EQR(\beta, r)\} = 1$$

(ii) for each fixed benefit-cost ratio $r > 1$

$$\liminf_{\beta \to 1} \sup\{\text{Eff}(\Pi_K) : \Pi_K \in EQR(\beta, r)\} = 1$$

In words: as agents become arbitrarily patient or the benefit/cost ratio becomes arbitrarily large, it is possible to choose *robust equilibrium protocols* that achieve efficiency arbitrarily close to first best. However, as the simulation illustrated in Figure 3 demonstrates, the protocols $\Pi_K$ *need not be the most efficient equilibrium protocols*; in this sense, $K/2$ is *not* the optimal quantity of money.[10]

Some intuition might be useful. Consider the protocols $\Pi_K$ and the corresponding invariant distributions. As $K$ increases, the fraction of agents who cannot purchase service and the fraction of agents who will not provide both decrease – so efficiency increases. However, if $r, \beta$ are fixed and $K$ increases then the protocols $\Pi_K$ will eventually cease to be equilibrium protocols so equilibrium efficiency is bounded. On the other hand, if we fix $r$ and let $\beta \to 1$ or fix $\beta$ and let $r \to \infty$ then the thresholds $K$ for which the protocols $\Pi_K$ *are* equilibrium protocols blow up, and hence efficiency tends to 1. Put differently: *high discount factors or high benefit/cost ratios make the use of high thresholds consistent with equilibrium.*

Theorem 9 provides asymptotic efficiency results; the following result presents an explicit lower bound (in terms of the population parameters $r, \beta$) for the efficiency obtainable by a robust equilibrium protocol.

---

[10]Contrast Berentsen (2000), for instance, in which $K/2$ is the optimal quantity of money – but only under the assumption that $K$ is an exogenous bound on token holdings.

**Theorem 10** *Given the benefit/cost ratio $r > 1$ and the discount factor $\beta < 1$, define*[11]

$$K^L = \max\left\{\log_{\frac{\rho\beta}{2(1-\beta)+2\rho\beta}}\left(\frac{1}{1+r}\right) - 1, 0\right\}$$

$$K^H = \log_{\frac{\rho\beta}{1-\beta+\rho\beta}}\left(\frac{1}{2r}\right)$$

*Then:*

(i) *all the thresholds $K$ for which $\Pi_K$ is a robust equilibrium protocol lie in the interval $[K^L, K^H]$;*

(ii) *the efficiency of the optimal robust equilibrium protocol is at least $\left(1 - \frac{1}{K^L+1}\right)^2 = \left(\frac{K^L}{K^L+1}\right)^2$.*

Theorem 10 yields a lower bound on efficiency because the optimal robust equilibrium protocol is at least as efficient as any protocol $\Pi_K$ that is a robust equilibrium, but does not yield an upper bound on efficiency because the optimal robust equilibrium protocol might be more efficient than any protocol $\Pi_K$ that is a robust equilibrium.

Theorem 10 also yields the designer an effective procedure for finding a robust equilibrium whose efficiency is *good*, if not optimal, since all that is necessary is to check protocols $\Pi_K$ with thresholds $K$ in the (finite) interval $[K^L, K^H]$. Moreover, it is not necessary to conduct an exhaustive search. Rather the designer can begin by checking the protocol $\Pi_K$, where $K$ is the midpoint of the interval $[K^L, K^H]$. If $M_{\sigma_K}(K-1) \geq c/\beta$ and $M_{\sigma_K}(K) \leq c/\beta$, then $\Pi_K$ is an equilibrium protocol and the search can stop. If $M_{\sigma_K}(K-1) < c/\beta$, then for all $K' > K$, $M_{\sigma_{K'}}(K'-1) < c/\beta$ (because $\beta^L(\Pi_{K'}) > \beta^L(\Pi_K)$). Therefore threshold protocols for which $K' > K$ cannot be an equilibrium and the designer can restrict search to the left half interval $[K^L, K]$. If $M_{\sigma_K}(K) > c/\beta$, then for all $K' < K$, $M_{\sigma_{K'}}(K') > c/\beta$ (because $\beta^H(\Pi_{K'}) > \beta^H(\Pi_K)$). Therefore threshold protocols for which $K' < K$ cannot be an equilibrium and the designer can restrict search to the right half interval $[K, K^H]$. Continuing to bisect in this way, the designer can find an equilibrium threshold protocol in at most $\log_2(K^H - K^L)$ iterations.

---

[11] Note that both the basis of the logarithms and the arguments are less than 1.

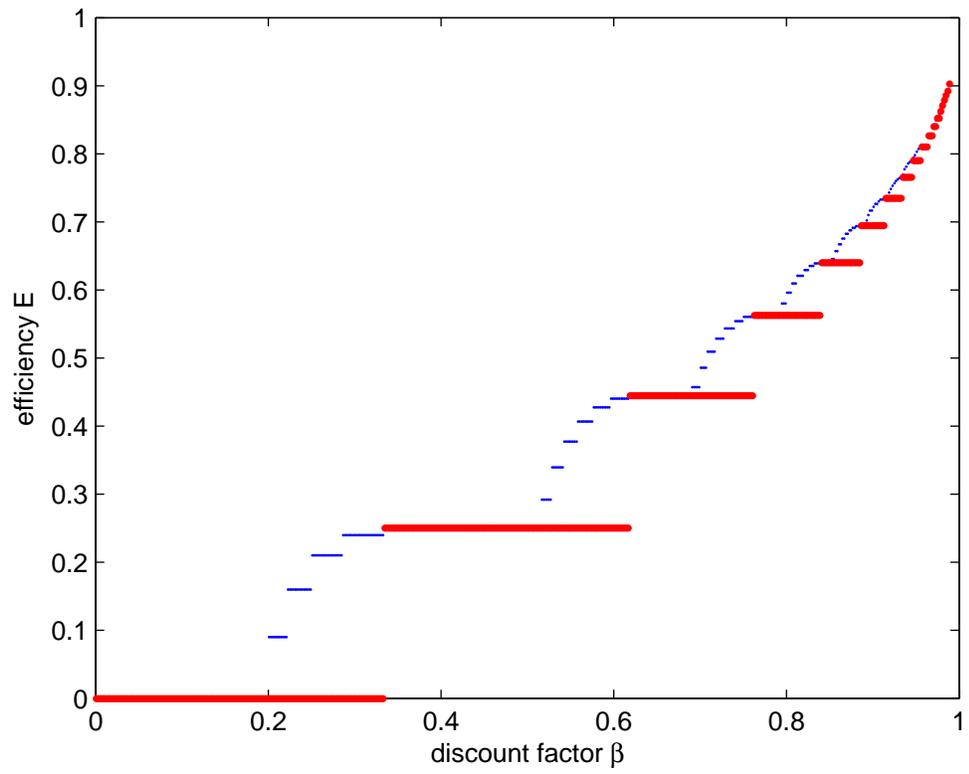

Figure 3: Optimal Equilibrium Protocols
(red (thick) - $\Pi_K$; blue (thin) - other protocols)

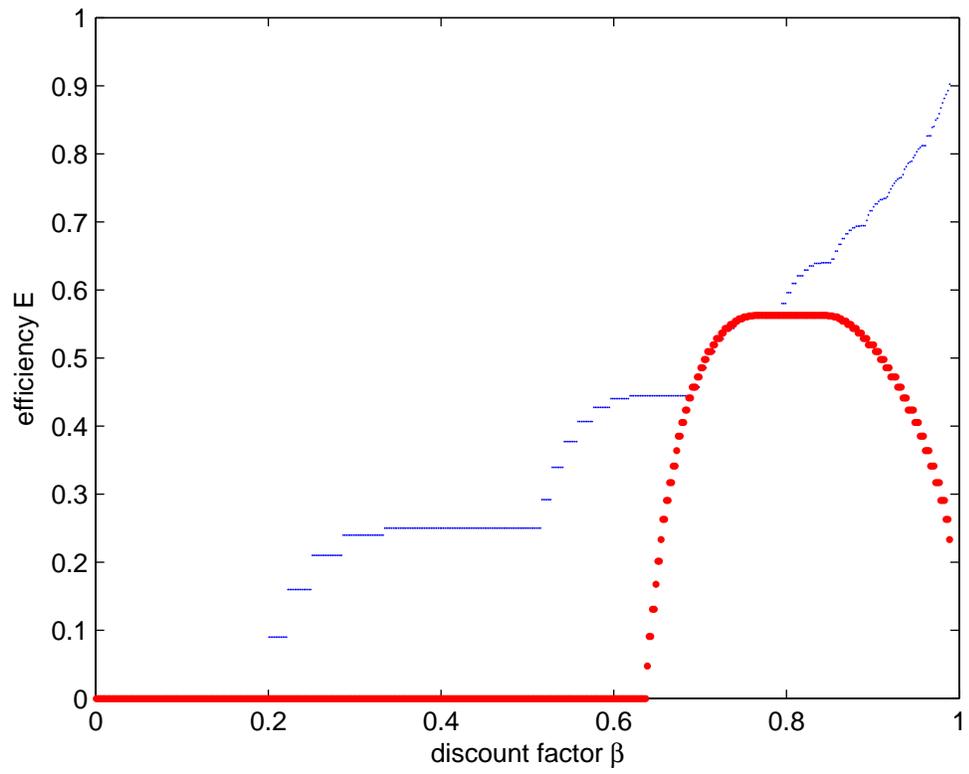

Figure 4: Inefficient Equilibrium Protocols
(red (thick) - $\Pi_3$; blue (thin) - optimal equilibrium protocols)

## 5 Choosing the Right Protocol

The reader may wonder why we have put so much emphasis on choosing the *right* protocol. As Figure 3 already shows, the reason is simple: choosing the wrong protocol can result in an enormous efficiency loss. Figure 4, which compares efficiency of the most efficient protocol with efficiency of a protocol for which the strategic threshold is constrained to be $K = 3$, makes this point in an even starker way: as the reader will see, except for a small range of discount factors, the efficiency loss is enormous.

# 6 Conclusion

In this paper, we have analyzed in some detail a simple, practicable and distributed method to incentivize trade in on-line environments through the use of (electronic) tokens. We have shown that when agents are patient, the method we offer can achieve outcomes that are nearly efficient, provided the right protocol (supply of tokens and recommended threshold) is chosen, but that equilibrium and efficiency are both sensitive to the precise choice of protocol. Surprisingly, the "optimal" supply of tokens need *not* be half the recommended threshold; this conclusion, and others, and much of the difficulty of our arguments are a consequence of our allowing agents to accumulate as many tokens as they wish, rather than imposing an exogenous bound on token holdings (which is common in the literature).

Our analysis is silent about convergence to the steady state. In particular, we do not know whether the recommended strategies would lead to convergence to the invariant distribution for all initial token distributions or for some particular token distributions. Berentsen (2000) proves convergence under some conditions, but in a continuous time model in which token holdings are subject to an exogenous bound. We have already noted that the latter is a strong (and, in our view, unrealistic) asumption, but another point is worth making. By definition, the recommended strategy is a best reply when the system is in the steady state, but the recommended strategy need not be a best reply — and very likely is *not* a best reply — when the system is *not* in the steady state — so why should agents follow it?

We have assumed that tokens are indivisible and must be exchanged for service one-for-one. Indivisibility seems an unavoidable consequence of the environment in which we are interested, but even if we were to allow for exchanges of fractional tokens, an equilibrium in which fractional tokens are exchanged cannot be robust, for the same reason that a mixed strategy equilibrium cannot be robust: any small perturbation of the population parameters will destroy the equilibrium. The effect of divisible money might be achieved by allowing for random offers and acceptances, as in Berentsen, Camera & Waller (2007), but again, equilibrium with random offers and acceptances could not be robust.

An alternative approach would be to allow agents to offer or demand any integral number of tokens, depending on their own holdings.[12] However, if agents can demand multiple tokens then it would seem that no matter an agent's current holding, he would *always* be willing to serve if the price of service was sufficiently high. Conversely, it would seem that no matter the price of service, an agent would be willing to pay if her token holding were sufficiently high. Hence, an equilibrium protocol in which agents could demand multiple tokens would seem to induce a stationary distribution with unbounded support and require offers and acceptances to depend on current holdings. It is not clear to us that such an equilibrium protocol could exist; certainly it would not seem to be practicable.[13]

We have considered the simplest setting, in which agents are identical, all files are equally valuable, and no errors occur. In a more realistic setting, we would need to take account of heterogeneous agents and files and allow for the possibilities of errors (in transmission of files or exchange of tokens or both). We have followed here the well-known adage "one has to start somewhere" – but we are keenly aware that there is much more work to be done.

---

[12]Much of the literature assumes that this is not possible, assuming instead that agents who "go looking" for trade can only carry a single token, no matter their current stock of tokens.

[13]Invariant distributions with unbounded support are impossible if the number of agents is finite, no matter how large; in this case it would also seem hard to see how a model with a continuum of agents would be a good proxy for a world with a finite, but very large, number of agents.

## Appendix: Proofs

**Proof of Theorem 1**  We first estimate $V(n+1) - V(n)$ (for $n \geq 0$) which is the loss from having one less token. To this end fix an optimal Markov strategy $(\tau, \sigma)$. We define a history-dependent strategy $(\tau', \sigma')$ and estimate the expected utility to an agent who begins with $n$ tokens and follows $(\tau', \sigma')$; this is a lower bound on $V(n)$. The strategy $(\tau', \sigma')$ is most easily described in the following way: Begin by following the behavior prescribed by the strategy $(\sigma, \tau)$ but for an agent who holds one more token than is actually held; i.e., $(\tau', \sigma')(h) = (\tau, \sigma)(N(h) + 1)$. If it never happens that the agent holds 0 tokens, requests service, and is matched with an agent who is willing to provide service, then continue in this way forever. If it does happen that the agent holds 0 tokens, requests service, and is matched with an agent who is willing to provide service, then service is not provided in that period (because the agent cannot pay) and after that period $(\tau', \sigma') = (\tau, \sigma)$. In other words, the agent behaves "as if" he held one more token than actually held until the first time such behavior results in requesting service, being offered service, and being unable to pay for service; after that point, revert to $(\tau, \sigma)$. The point to keep in mind is that if a moment of deviation occurs then an agent with one more token would hold exactly 1 token, would request and receive service, and in the next period would have 0 tokens – so that reverting to $(\tau, \sigma)$ is possible. Beginning with $n$ tokens and following the strategy $(\tau', \sigma')$ yields the same string of payoffs as beginning with $n+1$ tokens and following the strategy $(\tau, \sigma)$ except in the single period in which deviation occurs; in that period the expected loss of utility is at most $b\rho$. Hence the expected utility from beginning with $n$ tokens and following the strategy $(\tau', \sigma')$ yields utility at least $V(n+1) - b\rho$. Hence $V(n+1) - V(n) \leq b\rho < b < b/\beta$. However, this is the incentive compatibility condition that guarantees that an agent strictly prefers to request service when holding $n+1$ tokens, so the proof is complete. ∎

At this point it is convenient to collect some notation and isolate two technical results. Fix $\rho, b, c, \mu, \nu$ and consider a Markov strategy $\sigma$. For each $k$, let $V_\sigma(k, \beta)$ be the value of following $\sigma$ when the initial token holding is $k$ and the discount factor is $\beta$. As with the optimal value function $V$ defined in the text, the value function $V_\sigma$ can be defined by a recursive system of equations:

$$
\begin{aligned}
V_\sigma(0, \beta) &= \rho\sigma(0)[(1 - \mu)(-c + \beta V_\sigma(1, \beta)) \\
&\quad + \rho[1 - \sigma(0)]\beta V_\sigma(0, \beta) + (1 - 2\rho)\beta V_\sigma(0, \beta)
\end{aligned}
$$

$$
\begin{aligned}
V_\sigma(k, \beta) &= \rho[(1 - \nu)(b + \beta V_\sigma(k - 1, \beta) + \nu\beta V_\sigma(k, \beta)] \\
&\quad + \rho\sigma(k)[(1 - \mu)(-c + \beta V_\sigma(k + 1, \beta)) + \mu\beta V_\sigma(k, \beta)] \\
&\quad + \rho[1 - \sigma(k)]\beta V_\sigma(k, \beta) + (1 - 2\rho)\beta V_\sigma(k, \beta) \\
&\quad \text{for} \quad k > 0
\end{aligned}
\tag{9}
$$

From the value function, we define the marginal utilities

$$M_\sigma(k, \beta) = V_\sigma(k + 1, \beta) - V_\sigma(k, \beta) \tag{10}$$

If $\beta$ is fixed/understood, we simplify notation by writing $V_\sigma(k) = V_\sigma(k, \beta)$ and $M_\sigma(k) = M_\sigma(k, \beta)$.

It is also convenient to introduce some auxiliary parameters:

$$
\begin{aligned}
\phi_l &= -(1 - \nu)\rho\beta \\
\phi_c &= 1 - \beta + ((1 - \nu) + (1 - \mu))\rho\beta \\
\phi_r &= -(1 - \mu)\rho\beta
\end{aligned}
\tag{11}
$$

We note the signs of these parameters and various combinations:

$$
\begin{array}{ccc}
\phi_l < 0, & \phi_c > 0, & \phi_r < 0 \\
\phi_l + \phi_c + \phi_r > 0, & \phi_l + \phi_c > 0, & \phi_r + \phi_c > 0
\end{array}
\tag{12}
$$

Using these auxiliary parameters and the recursion relations for $V(\sigma$ and performing some simple algebraic manipulations yields a useful matrix representation involving marginals that we will use frequently:

$$
\begin{bmatrix}
\phi_c & \phi_r & 0 & \cdots & 0 \\
\phi_l & \phi_c & \phi_r & 0 & \vdots \\
0 & \phi_l & \phi_c & \phi_r & 0 \\
\vdots & \ddots & \ddots & \ddots & \ddots \\
0 & \cdots & 0 & \phi_l & \phi_c
\end{bmatrix}_{K \times K}
\begin{bmatrix}
M_\sigma(0) \\
M_\sigma(1) \\
\vdots \\
M_\sigma(K_1 - 1)
\end{bmatrix}
=
\begin{bmatrix}
(1 - \nu)\rho b \\
0 \\
\vdots \\
0 \\
(1 - \mu)\rho c
\end{bmatrix}
\tag{13}
$$

In short form, write this matrix representation as

$$\Phi \mathbf{M} = \mathbf{u} \tag{14}$$

**Lemma 1** *Fix $\rho, b, c, \mu, \nu$ and $\beta$. Let $\sigma$ be a Markov strategy with the property that*

$$\sigma(k) = \begin{cases} 1 & \text{if} \quad 0 \leq k < K_1 \\ 0 & \text{if} \quad K_1 \leq k < K_2 \end{cases}$$

*Then:*

(i) *if $0 \leq k < K_2$ then $M_\sigma(k) > 0$*

(ii) *in the range $0 \leq k < K_1$, $M_\sigma$ is either increasing, decreasing or decreasing then increasing*

(iii) *if $M_\sigma(K_1-1) \geq c/\beta$ then $M_\sigma(0) > M_\sigma(1) > ...M_\sigma(K_1-2) \geq M_\sigma(K_1-1) \geq c/\beta$*

**Proof** We first consider the token holding levels $0 \leq k < K_1$. We make use of the matrix representation (13).

To prove (i), we first show that $M_\sigma(k) > 0$ for $0 \leq k < K_1$. If $K_1 < 3$, this follows by simply solving the matrix representation, so we henceforward assume $K_1 \geq 3$. If there exists a token holding level $k^*$ with $0 \leq k^* < K_1$ such that $M_\sigma(k^*) \leq 0$ then one of the following must hold: either (a) there two consecutive such token holding levels, or (b) the marginal payoffs of the neighboring token holding levels are both positive. We consider these cases separately.

(a) In this case, there exists $k^*$ such that $M_\sigma(k^*), M_\sigma(k^*+1)$ are both non-positive. Of these marginals, one is at least as big; say $M_\sigma(k^*) \geq M_\sigma(k^*+1)$. From the identities above we see that

$$\begin{aligned} M_\sigma(k^*+2) &= \frac{\phi_l M_\sigma(k^*) + \phi_c M_\sigma(k^*+1)}{-\phi_r} \\ &\leq \frac{(\phi_l + \phi_c)M_\sigma(k^*+1)}{-\phi_r} \\ &\leq M_\sigma(k^*+1) \end{aligned}$$

Proceeding inductively, it follows that $0 \geq M_\sigma(k^*) \geq M_\sigma(k^* + 1)... \geq M_\sigma(K_1 - 1)$. Moreover,

$$\begin{aligned}\phi_c M_\sigma(K_1 - 1) &= (1 - \mu)\rho c - \phi_l M_\sigma(K_1 - 2) \\ &> -\phi_l M_\sigma(K_1 - 2) \\ &> -\phi_l M(K - 1)\end{aligned}$$

This requires $\phi_c < -\phi_l$ which contradicts the sign relations (12). The argument when $M_\sigma(k^*+1) \geq M_\sigma(k^*)$ is similar and is left for the reader.

(b) In this case, there exists $k^*$ such that $M_\sigma(k^* - 1) > 0$, $M_\sigma(k^*) \leq 0$, $M_\sigma(k^* + 1) > 0$. This entails

$$\phi_l M_\sigma(k^* - 1) + \phi_c M_\sigma(k^*) + \phi_r M_\sigma(k^* + 1) < 0 \tag{15}$$

which again contradicts the sign relations (12).

From the above we conclude $M_\sigma(k) > 0$ for $0 < k < K_1$. To see that $M_\sigma(0) > 0$ note that

$$-\phi_r M_\sigma(1) = \phi_c M_\sigma(0) - (1 - \nu)\rho b < -\phi_r M_\sigma(0) \tag{16}$$

Therefore, $M_\sigma(1) < M_\sigma(0)$, so $M_\sigma(0) > 0$, as desired.

Finally, to see that $M_\sigma(k) > 0$ for $K_1 \leq k < K_2$, apply the recursion equations (9) to obtain

$$\phi_l M_\sigma(k - 1) + (\phi_c + \phi_r)M_\sigma(k) = 0 \tag{17}$$

We know that $M_\sigma(K_1-1) > 0$ so the sign relations (12) imply that $M_\sigma(K_1) > 0$ as well. Now it follows inductively that $M_\sigma(k) > 0$ for $K_1 \leq k < K_2$. This completes the proof of (i).

To prove (ii) it is enough to show that $M_\sigma$ has no local maximum for $0 < k < K_1$. If $M$ had a local maximum $k^*$ in this range we would have $M_\sigma(k^*) \geq M_\sigma(k^* - 1)$ and $M_\sigma(k^*) \geq M_\sigma(k^* + 1)$. However, algebraic manipulation yields the inequalities

$$\begin{aligned}M_\sigma(k^*) &= \frac{-\phi_l M_\sigma(k^* - 1) - \phi_r M_\sigma(k^* + 1)}{\phi_c} \\ &\leq \frac{-\phi_l - \phi_r}{\phi_c} M_\sigma(k^*) \\ &< M_\sigma(k^*)\end{aligned}$$

which is a contradiction. This establishes (ii)

To prove (iii), first manipulate the matrix identity (13) to obtain:

$$\begin{aligned} &(1-\nu)\rho\beta M_\sigma(K_1-2) \\ &= (1-\beta+((1-\nu)+(1-\mu))\rho\beta)M_\sigma(K-1) - (1-\mu)\rho c \\ &\geq (1-\beta+(1-\nu)\rho\beta)M_\sigma(K_1-1) \geq (1-\nu)\rho\beta M_\sigma(K_1-1) \end{aligned} \quad (18)$$

In view of (ii), the marginal payoffs are decreasing, so this establishes (iii). ∎

**Lemma 2** *Fix $\rho, b, c$ and a threshold protocol $\Pi = (\alpha, \sigma_K)$ with corresponding $\mu^\Pi, \nu^\Pi$. The marginal utility $M_{\sigma_K}(k, \beta)$ is strictly increasing in the discount factor $\beta$, i.e., if $0 \leq \beta_1 < \beta_2 < 1$, then,*

$$M_{\sigma_K}(k, \beta_1) < M_{\sigma_K}(k, \beta_2) \quad \text{for all } k \quad (19)$$

**Proof** To economize slightly on notation we write $\sigma = \sigma_K$. We present the proof in three steps.

In Step 1, we prove that if there exist $0 < K_1 \leq K_2 < K - 1$ such that $\forall k \in [K_1, K_2], M_\sigma(k, \beta_1) \geq M_\sigma(k, \beta_2)$, then at least one of the following is true, $M_\sigma(K_1 - 1, \beta_1) \geq M_\sigma(K_1 - 1, \beta_2)$ or $M_\sigma(K_2 + 1, \beta_1) \geq M_\sigma(K_2 + 1, \beta_2)$.

In Step 2, we prove that if there exists a $k^* \in [0, K-1]$ such that $M_\sigma(k^*, \beta_1) \geq M_\sigma(k^*, \beta_2)$, then for all $k \in [0, K-1], M_\sigma(k, \beta_1) \geq M_\sigma(k, \beta_2)$. Step 2 uses the result of Step 1.

In Step 3, we disprove the possibility that $k \in [0, K-1], M_\sigma(k, \beta_1) \geq M_\sigma(k, \beta_2)$.

Step 2 and Step 3 together show a contradiction and therefore, $k \in [0, K-1], M_\sigma(k, \beta_1) < M_\sigma(k, \beta_2)$.

**Step 1** We assert that if there are indices $0 < K_1 \leq K_2 < K - 1$ such that $M_\sigma(k, \beta_1) \geq M_\sigma(k, \beta_2)$ for all $K_1 \leq k \leq K_2$ then at least one of the following must hold:

(A) $M_\sigma(K_1 - 1, \beta_1) \geq M_\sigma(K_1 - 1, \beta_2)$

(B) or $M_\sigma(K_2 + 1, \beta_1) \geq M_\sigma(K_2 + 1, \beta_2)$.

To see this, note that simple manipulations of the matrix representation (13) yield

- if $K_2 = K_1$ then

$$(1-\nu)\rho M_\sigma(K_1-1,\beta) + (1-\mu)\rho M_\sigma(K_2+1,\beta)$$
$$= (1/\beta - 1 + ((1-\nu)+(1-\mu))\rho)M_\sigma(K_1,\beta)$$

- if $K_2 > K_1$ then

$$(1-\nu)\rho M_\sigma(K_1-1,\beta) + (1-\mu)\rho M_\sigma(K_2+1,\beta)$$
$$= (1/\beta - 1 + (1-\mu)\rho)M_\sigma(K_1,\beta)$$
$$= +(1/\beta - 1)[M_\sigma(K_1+1,\beta) + ... + M_\sigma(K_2-1,\beta)]$$
$$= +(1/\beta - 1 + (1-\nu)\rho)M_\sigma(K_2,\beta)$$

Since $\beta_1 < \beta_2$ and we have assumed $M_\sigma(k,\beta_1) \geq M_\sigma(k,\beta_2)$ for $0 < K_1 \leq K_2 < K-1$, in each of the cases above the right-hand side is larger when $\beta = \beta_1$ than when $\beta = \beta_2$. Because the terms in the left-hand sides are positive, it follows that at least one of (A), (B) must hold, as asserted.

**Step 2** We assert first that if there is a $k^*, 0 \leq k^* \leq K_1$ such that $M_\sigma(k^*,\beta_1) \geq M_\sigma(k^*,\beta_2)$, then at least one of the following must hold:

(C) there exists some $K_3$, $0 \leq K_3 \leq K_1$, such that $M_\sigma(k,\beta_1) \geq M_\sigma(k,\beta_2)$ for all $k$, $0 \leq k \leq K_3$

(C) there exists some $K_4$, $0 \leq K_4 \leq K_1$, such that $M_\sigma(k,\beta_1) \geq M_\sigma(k,\beta_2)$ for all $k$, $K_4 \leq k \leq K_-1$

To see this, note first that if $k^* = 0$ satisfies the hypothesis, then (C) holds with $K_3 = 0$ and that if $k^* = K-1$ satisfies the hypothesis, then (D) holds with $K_4 = K-1$. Hence it suffices to consider a $k^*$, $0 < k^* < K-1$, that satisfies the hypothesis. We now make use of Step 1. Set $K_1 = K_2 = k^*$. Applying Step 1 once increases the token holding interval where $M_\sigma(k,\beta_1) \geq M_\sigma(k,\beta_2)$ by 1. Let $K_1$ and $K_2$ be the new end points of the interval and apply Step 1 again. Continuing in this way we come eventually to a point where either $K_1 = 0$ or $K_2 = K-1$. If $K_1 = 0$, set $K_3 = K_2$ and note that (C) holds. If $K_2 = K-1$, set $K_4 = K-1$ and note that (D) holds

We now show that either (C) or (D) leads to the desired conclusion. Consider (C) first. Using the matrix representation (13) we obtain

$$
\begin{aligned}
(1-\nu)\rho\beta M_\sigma(K_1+1,\beta) \; + \; & (1-\nu)\rho b \\
= \; & [1-(1-(1-\mu)\rho)\beta]M_\sigma(0,\beta) \\
& + (1-\beta)[M_\sigma(1,\beta)+...+M_({K_1-1,\beta)}] \\
& + [1-(1-(1-\nu)\rho)\beta]M_\sigma(K_1,\beta)
\end{aligned}
$$

The right-hand side is bigger when $\beta = \beta_1$ than when $\beta = \beta_2$. Therefore $M_\sigma(K_1+1,\beta_1) \geq M_\sigma(K_1+1,\beta_2)$. By induction, $M_\sigma(k,\beta_1) \geq M_\sigma(k,\beta_2)$ for all $k$, $0 \leq k \leq K-1$.

Now consider (D). Using the matrix representation (13) we obtain

$$
\begin{aligned}
(1-\mu)\rho\beta M_\sigma(K_2-1,\beta) \; + \; & (1-\mu)\rho c \\
= \; & [1-(1-(1-\nu)\rho)\beta]M_\sigma(K-1,\beta) \\
& + (1-\beta)[M_\sigma(K-2,\beta)+...+M_\sigma(K_2+1,\beta)] \\
& + [1-(1-(1-\mu)\rho)\beta]M_\sigma(K_2,\beta)
\end{aligned}
$$

The right-hand side is bigger when $\beta = \beta_1$ than when $\beta = \beta_2$. Therefore $M_\sigma(K_2-1,\beta_1) \geq M_\sigma(K_2-1,\beta_2)$. By induction, $M_\sigma(k,\beta_1) \geq M_\sigma(k,\beta_2)$ for all $k$, $0 \leq k \leq K-1$.

Taking (C) and (D) together completes Step 2.

**Step 3** Using the matrix representation (13) we obtain

$$
\begin{aligned}
& [1-(1-(1-\mu)\rho)\beta]M_\sigma(0,\beta) \\
& +(1-\beta)[M_\sigma(1,\beta)+...+M_\sigma(K_1-1,\beta)] \\
& +[1-(1-(1-\nu)\rho)\beta]M_\sigma(K-1,\beta) \\
= \; & (1-\nu)\rho b + (1-\mu)\rho c
\end{aligned}
$$

In view of Step 2, the left-hand side is bigger when $\beta = \beta_1$ than when $\beta = \beta_2$. However, the right-hand side is independent of $\beta$, so this is a contradiction. We conclude that $M_\sigma(k,\beta_1) < M_\sigma(k,\beta_2)$ for every $k$, $0 \leq k \leq K-1$. ∎

**Proof of Theorem 2** Fix $\beta$. The Markov strategy $\sigma$ is optimal if and only if it satisfies the Bellman optimality conditions:

$$\beta(V_\sigma(k+1) - V_\sigma(k)) \geq c, \text{if } \sigma(k) = 1 \qquad (20)$$
$$\beta(V_\sigma(k+1) - V_\sigma(k)) \leq c, \text{if } \sigma(k) = 0 \qquad (21)$$

If $\sigma$ is not a threshold strategy, there must exist integers $K_1 < K_2$ such that

$$\begin{aligned} \sigma(k) &= 1, & 0 \leq k < K_1 \\ \sigma(k) &= 0, & K_1 \leq k < K_2 \\ \sigma(k) &= 1, & k = K_2 \end{aligned} \qquad (22)$$

We will show that the Bellman optimality conditions are violated at $K_2$ and $K_2 - 1$. To this end, let $K_3$ be the smallest integer greater than $K_2$ for which $\sigma(K_3) = 0$. (Such an integer exists because it cannot be optimal to serve when the token holding is sufficiently high.) Thus $\sigma(k) = 1$, for $K_2 \leq k < K_3$ and $M_\sigma(K_3 - 1) \geq c/\beta$. Following $\sigma$,

$$M_\sigma(K_3 - 2) = [(1-\mu)\rho c - \phi_c M_\sigma(K_3 - 1)]/\phi_l > M_\sigma(K_3 - 1) \geq c/\beta \qquad (23)$$

An inductive argument shows that $M_\sigma(K_2) > M_\sigma(K_2+1) \geq c/\beta$. According to the recursion equations (9) we have

$$M_\sigma(K_2 - 1) = (\phi_c M_\sigma(K_2) + \phi_r M_\sigma(K_2 + 1))/(-\phi_l) > c/\beta$$

which is a contradiction. We conclude that a non-threshold strategy cannot be optimal; equivalently, only threshold strategies can be optimal strategies.

It remains to show that the only possible optimal threshold strategies have adjacent thresholds. Consider first two threshold strategies with consecutive thresholds $K$ and $K + 1$. We assert that

$$M_{\sigma_K}(K) < c/\beta \Leftrightarrow M_{\sigma_{K+1}}(K) < c/\beta \qquad (24)$$

We prove direction "$\Rightarrow$"; the "$\Leftarrow$" direction is similar and left to the reader. Suppose instead that $M_{\sigma_{K+1}}(K) \geq c/\beta$. It follows that $-\phi_r M_{\sigma_{K+1}}(K) \geq (1-\mu)\rho c$. If we delete the last line in the matrix equation (13) for $\sigma_{K+1}$ and move $M_{\sigma_{K+1}}(K)$ to the right-hand side, we get another matrix equation

$$\Phi_{K \times K} \mathbf{M}_{\sigma_{K+1}} = \tilde{\mathbf{u}}$$

where $\tilde{\mathbf{u}} = ((1-\nu)\rho b, 0, ..., 0, -\phi_r M_{\sigma_{K+1}}(K))^{\mathrm{T}}$. For the threshold $K$, $\Phi_{K \times K} \mathbf{M}_{\sigma_K} = \mathbf{u}$. Therefore,

$$\Phi_{K \times K}(\mathbf{M}_{\sigma_{K+1}} - \mathbf{M}_{\sigma_K}) = \tilde{\mathbf{u}} - \mathbf{u} \tag{25}$$

Lemma 1 guarantees that $\tilde{\mathbf{u}} - \mathbf{u} \geq 0$, so $\mathbf{M}_{\sigma_{K+1}} \geq \mathbf{M}_{\sigma_K}$. That is, $M_{\sigma_{K+1}}(k) \geq M_{\sigma_K}(k)$ for $0 \leq k \leq K-1$. Because $M_{\sigma_{K+1}}(K) \geq c/\beta > M_{\sigma_K}(K)$, it follows that $M_{\sigma_{K+1}}(k) \geq M_{\sigma_K}(k)$ for $0 \leq k \leq K$. According to the matrix equation, the following identity holds for both $\sigma = \sigma_K$ and $\sigma = \sigma_{K+1}$:

$$\begin{aligned}
&(1-\nu)\rho b + (1-\mu)\rho c \\
=& (1 - \beta + (1-\mu)\rho\beta) M_\sigma(0) \\
&+ (1-\beta) \sum_{k=1}^{K-1} M_\sigma(k) + (1 - \beta + (1-\nu)\rho\beta) M_\sigma(K)
\end{aligned} \tag{26}$$

This is a contradiction so we have established the direction $\Rightarrow$, as desired.

It follows directly from the matrix identity that

$$M_{\sigma_K}(K) = c/\beta \Leftrightarrow M_{\sigma_{K+1}}(K) = c/\beta$$

Hence

$$M_{\sigma_K}(K) > c/\beta \Leftrightarrow M_{\sigma_{K+1}}(K) > c/\beta \tag{27}$$

We now assert that if $\tilde{K} > K$ then

$$M_{\sigma_K}(K) < c/\beta \Rightarrow M_{\sigma_{\tilde{K}}}(\tilde{K} - 1) < c/\beta \tag{28}$$

We have already shown that this is true when $\tilde{K} = K+1$; i.e. $M_{\sigma_{K+1}}(K) < c/\beta$. Consider $\tilde{K} = K+2$. Of $M_{\sigma_{K+2}}(K+1) \geq c/\beta$, then (27) implies that $M_{\sigma_{K+1}}(K+1) \geq c/\beta$. Therefore, $M_{\sigma_{K+1}}(K+1) > M_{\sigma_{K+1}}(K)$. This is a contradiction to $M_{\sigma_{K+1}}(K+1) < M_{\sigma_{K+1}}(K)$. Following inductively we obtain the assertion (28).

A similar argument (which we omit) shows that:

$$M_{\sigma_K}(K-1) > c/\beta \Rightarrow M_{\sigma_{\tilde{K}}}(\tilde{K}) > c/\beta, \forall \tilde{K} < K \tag{29}$$

Finally, suppose $\sigma_K$ is an optimal threshold strategy. Then $M_{\sigma_K}(K-1) \geq c/\beta$ and $M_{\sigma_K}(K) \leq c/\beta$. If the equalities hold strictly, (28) and (29) guarantee that $\sigma_K$ is the only optimal threshold strategy. If $M_{\sigma_K}(K-1) = c/\beta$ (and hence, $M_{\sigma_K}(K) < c/\beta$), only $\sigma_K$ and $\sigma_{K-1}$ are optimal threshold strategies. If $M_{\sigma_K}(K) = c/\beta$ (and hence, $M_{\sigma_K}(K-1) > c/\beta$), only $\sigma_K$ and $\sigma_{K+1}$ are optimal threshold strategies. This completes the proof. ∎

**Proof of Theorem 3** This follows immediately from the representation of $\eta_+$ and the definition of invariance . ∎

**Proof of Theorem 4** Given a protocol $\Pi = (\alpha, \sigma)$, let $\eta^\Pi$ be the unique invariant distribution; let $\mu^\Pi$ be the fraction of agents who have no tokens and $\nu^\Pi$ the fraction of agents who do not provide service; these depend only on $\Pi$ and not on the population parameters. If $\sigma = \sum \gamma(K)\sigma_K$ is a best response given the population parameters and $\mu^\Pi, \nu^\Pi$, $\gamma$ must put strictly positive weight only on threshold strategies $\sigma_K$ that are pure best responses. In view of Theorem 2, there are at most two threshold strategies that are pure best responses and they are at adjacent thresholds. That is, $\sigma$ is either a pure threshold strategy or a mixture of two adjacent threshold strategies, as asserted. ∎

**Proof of Theorem 5** Suppose to the contrary that $\Pi = (\alpha, \sigma)$ is a robust equilibrium protocol and that $\sigma = \sum \gamma(K)\sigma_K$ is a proper mixed strategy, so that $\gamma(K) > 0$ for at least two values of the threshold $K$, Let $\mu^\Pi$ be the fraction of agents who have no tokens and $\nu^\Pi$ the fraction of agents who do not provide service; these depend only on $\Pi$ and not on the population parameters. In view of Theorem 4, $\sigma$ must assign positive probability only to two adjacent threshold strategies; say $\sigma = \gamma(K)\sigma_K + \gamma(K+1)\sigma_{K+1}$ with $\gamma(K) > 0$ and $\gamma(K+1) > 0$, and both $\sigma_K, \sigma_{K+1}$ must be best responses. Because $\sigma_K(K+1) = 0$ and $\sigma_{K+1}(K+1) = 1$, equations (8), (9) (which provide necessary and sufficient conditions for optimality in terms of the *true* value function) entail that

$$-c + \beta V_{K+1} \leq \beta V_K$$
$$-c + \beta V_{K+1} \geq \beta V_K$$

Hence $-c + \beta V_{K+1} = \beta V_K$. Because $\sigma_K$ is a best response, the value functions $V_{\sigma_K}$ must coincide with the true value function $V$. Hence, an agent following $\sigma_K$ must be indifferent to providing service when holding $K$ tokens. However, if $\beta$ increases slightly $M_{\sigma_K}$ also increases, whence an agent following $\sigma_K$ must strictly prefer to provide service. In other words, when $\beta$ increases slightly, $\sigma_K$ can no longer be a best response and $\sigma_K$ can no longer be an equilibrium protocol. This is a contradiction, so we conclude that a robust equilibrium protocol $\Pi$ cannot involve proper mixed strategies, as asserted. ∎

**Proof of Theorem 6** We divide the proof of (i) into several steps.

**Step 1** We first prove there exists $\beta^L \in [0, 1)$ such that

$$\begin{aligned} M_\sigma(K-1, \beta) &< \frac{c}{\beta} \text{ for } \beta < \beta^L \\ M_\sigma(K-1, \beta^L) &= \frac{c}{\beta} \\ M_\sigma(K-1, \beta) &> \frac{c}{\beta} \text{ for } \beta > \beta^L \end{aligned}$$

To see this, define the auxiliary function

$$F(\beta) = M_\sigma(K-1, \beta) - \frac{c}{\beta}$$

$F$ is evidently continuous. Lemma 2 guarantees that $M_\sigma(K-1, \beta)$ is strictly increasing in $\beta$, so $F(\beta)$ is also strictly increasing in $\beta$ as well. We show that $F(1) > 0$ and $\lim_{\beta \to 0} F(\beta) < 0$ and then apply the intermediate value theorem to find $\beta^L$.

To see that $F(1) > 0$, note first that the coefficients in the left-hand matrix of (13) are simply $\phi_l = -\rho(1-\nu)$, $\phi_c = \rho(1-\nu+1-\mu)$ and $\phi_r = \rho(1-\mu)$. We split the matrix $\mathbf{M}_{\sigma_K}$ in two parts. To do this, write

$$\begin{aligned} \mathbf{u}' &= (\rho(1-\nu)c \quad 0 \quad \ldots \quad 0 \quad \rho(1-\mu)c)^T \\ \mathbf{u}'' &= (\rho(1-\nu)(b-c) \quad 0 \quad \ldots \quad 0 \quad 0)^T \end{aligned} \quad (30)$$

and define $\mathbf{M}'_{\sigma_K}, \mathbf{M}''_{\sigma_K}$ to be the solutions to the equations

$$\Phi \mathbf{M}'_{\sigma_K} = \mathbf{u}', \quad \Phi \mathbf{M}''_{\sigma_K} = \mathbf{u}'' \quad (31)$$

Note that $\mathbf{M}_{\sigma_K} = \mathbf{M}'_{\sigma_K} + \mathbf{M}''_{\sigma_K}$ and $\mathbf{M}_{\sigma_K}$ is the solution to (13). It is easy to check that $\mathbf{M}'_{\sigma_K}$ is a constant matrix: $M'_{\sigma_K}(k) = c$ for $0 \leq k < K-1$.

Lemma 1 guarantees that the entries of $\mathbf{M}''_{\sigma_K}$ are strictly positive: $M''_\sigma(k) > 0$ for $0 \leq k < K-1$. Hence the entries of $\mathbf{M}_{\sigma_K}$ are strictly greater than $c$: $M_\sigma(k) > c$ for $0 \leq k < K-1$. In particular, $F(1) > 0$.

To see that $\lim_{\beta \to 0} F(\beta) < 0$, suppose not. Because $F$ is strictly increasing, this means $F(\beta) \geq 0$ for every $\beta \in (0,1]$, which entails that $M_\sigma(k) \geq \frac{c}{\beta}$ for $0 \leq k < K-1$. Summing the rows in (13) yields:

$$\rho(1-\nu)b + \rho(1-\mu)c > K(1-\beta)\frac{c}{\beta} = \frac{Kc}{\beta} - Kc \tag{32}$$

Note that $Kc/\beta$ flows up as $\beta \to 0$, so this is impossible. We conclude that $\lim_{\beta \to 0} F(\beta) < 0$, as asserted.

Because $F$ is strictly increasing, the intermediate value theorem guarantees that we can find an unique $\beta^L$ such that

$$\begin{array}{rcl} F(\beta) & < & 0 \quad \text{for} \quad \beta < \beta^L \\ F(\beta^L) & = & 0 \\ F(\beta) & > & 0 \quad \text{for} \quad \beta > \beta^L \end{array}$$

The definition of $F$ yields the desired property of $\beta^L$

**Step 2** Next we prove there exists $\beta^H \in (\beta^L, 1)$ such that if $\beta \in [0, \beta^H]$ then

$$\begin{array}{rcl} M_{\sigma_K,\beta}(K-1) & < & \frac{\phi_c + \phi_r}{-\phi_l}\frac{c}{\beta} \quad \text{for } \beta < \beta^H \\ M_{\sigma_K,\beta^H}(K-1) & = & \frac{\phi_c + \phi_r}{-\phi_l}\frac{c}{\beta} \\ M_{\sigma_K,\beta}(K-1) & > & \frac{\phi_c + \phi_r}{-\phi_l}\frac{c}{\beta} \quad \text{for } \beta > \beta^H \end{array}$$

To see this, note first that $\frac{\phi_c + \phi_r}{-\phi_l}\frac{c}{\beta} = \left[1 - \frac{1}{\rho(1-\nu)} + \frac{1}{\rho(1-\nu)\beta}\right]\frac{c}{\beta}$ and define another auxiliary function:

$$G(\beta) = M_\Pi(K-1, \beta) - (1 - \frac{1}{\rho(1-\nu)} + \frac{1}{\rho(1-\nu)\beta})\frac{c}{\beta}$$

$G$ is continuous and increasing. From Step 1 it follows that $M_{\sigma_K}(K-1,1) > c$ so $G(1) = M_{\sigma_K}(K-1,1) - c > 0$. It also follows that $M_{\sigma_K}(K-1, \beta^L) = \frac{c}{\beta^L}$; because $(1 - \frac{1}{\rho(1-\nu)} + \frac{1}{\rho(1-\nu)\beta^L})\frac{c}{\beta^L} > \frac{c}{\beta^L}$, we conclude that $G(\beta^L) < 0$.

Because $G$ is continuous and increasing, there is a unique $\beta^H \in (\beta^L, 1)$ such that
$$\begin{aligned} G(\beta) &< 0 \quad \text{for} \quad \beta < \beta^H \\ G(\beta^H) &= 0 \\ G(\beta) &> 0 \quad \text{for} \quad \beta > \beta^L \end{aligned}$$

**Step 3** The definitions of $F, G$ imply that in order for $\Pi$ to be an equilibrium protocol when the discount factor is $\beta$ it is the necessary and sufficient condition that $F(\beta) \geq 0$ and $G(\beta) \leq 0$. Hence $\Pi$ is an equilibrium protocol when the discount factor is $\beta$ exactly for $\beta \in [\beta^L, \beta^H]$.

Because $F, G$ are continuous in all their arguments and strictly increasing, $\beta^L, \beta^H$, which are the zeroes of $F, G$, are continuous functions of the parameters as well. This completes the proof of (i).

The proof of (ii) is similar and left to the reader. ∎

**Proof of Theorem 7** We first consider (i). Fix $r$. Consider the two protocols $\Pi_K = (K/2, \sigma_K)$ and $\Pi_{K+1} = ((K+1)/2, \sigma_{K+1})$ and the corresponding intervals $[\beta_1^L, \beta_1^H]$ and $[\beta_2^L, \beta_2^H]$ of discount factors that sustain equilibrium. We need to show that
$$\beta_1^L < \beta_2^L < \beta_1^H < \beta_2^H$$

(The sustainable ranges overlap between two consecutive threshold protocols overlap but are not nested.) There are three inequalities to be established; we carry out the analyses in (A), (B), (C) below.

There are

(A) To prove $\beta_2^L > \beta_1^L$, write $\beta = \beta_1^L$. We show that $M_{\sigma_{K+1}}(K) < \frac{c}{\beta}$. To see this, suppose not; i.e. $M_{\sigma_{K+1}}(K) \geq \frac{c}{\beta}$. The construction of $\beta_1^L$ guarantees that $M_{\sigma_K}(K-1) = c/\beta$. We will use this inequality and equality to show that *all* marginal payoffs of $\Pi_{K+1}$ so large that they violate the restrictions imposed by the bounded benefit $b$ and cost $c$.

To simplify the notation, let $\omega_X = \frac{X+1}{X}(\frac{1}{\beta} - 1)\frac{1}{\rho}$. Note $\omega_{K+1} < \omega_K$. Then the matrix identity (13) becomes:

$$\begin{bmatrix} \omega_X+2 & -1 & 0 & \cdots & 0 \\ -1 & \omega_X+2 & -1 & 0 & \vdots \\ 0 & -1 & \omega_X+2 & -1 & 0 \\ \vdots & \ddots & \ddots & \ddots & \ddots \\ 0 & \cdots & 0 & -1 & \omega_X+2 \end{bmatrix}_{X \times X} \begin{bmatrix} M_{\sigma_X}(0) \\ M_{\sigma_X}(1) \\ \vdots \\ M_{\sigma_X}(X-1) \end{bmatrix} = \begin{bmatrix} b/\beta \\ 0 \\ \vdots \\ 0 \\ c/\beta \end{bmatrix}$$

(33)

Suppose $M_{\sigma_{K+1}}(K) \geq M_{\sigma_K}(K-1) = \frac{c}{\beta}$. We investigate the relation between $M_{\sigma_{K+1}}(K-1)$ and $M_{\sigma_K}(K-2)$. Using the matrix identity,

$$\frac{M_{\sigma_{K+1}}(K-1)}{M_{\sigma_K}(K-2)} = \frac{(\omega_{K+1}+2)M_{\sigma_{K+1}}(K) - \frac{c}{\beta}}{(\omega_K+2)M_{\sigma_K}(K-1) - \frac{c}{\beta}}$$
$$> \frac{(\omega_{K+1}+2)M_{\sigma_{K+1}}(K)}{(\omega_K+2)M_{\sigma_K}(K-1)} > \frac{\omega_{K+1}+1}{\omega_K+1}$$

Moreover if $2 \leq k \leq K-1$ then

$$\frac{M_{\sigma_{K+1}}(K-k)}{M_{\sigma_K}(K-k-1)} = \frac{(\omega_{K+1}+1)[M_{\sigma_{K+1}}(K) + M_{\sigma_{K+1}}(K-k+1)] - \frac{c}{\beta}}{(\omega_K+1)[M_{\sigma_K}(K-1) + M_{\sigma_K}(K-k)] - \frac{c}{\beta}}$$

By induction,

$$\frac{M_{\sigma_{K+1}}(K-k)}{M_{\sigma_K}(K-k-1)} > \left(\frac{\omega_{K+1}+1}{\omega_K+1}\right)^k > \left(\frac{\omega_{K+1}}{\omega_K}\right)^k > \left(1 - \frac{1}{(K+1)^2}\right)^k$$
$$> 1 - \frac{k}{(K+1)^2} > \frac{K+1}{K+2}, \forall 0 \leq k \leq K-1$$

Next we prove $M_{\sigma_{K+1}}(0) \geq M_{\sigma_K}(0)$. This is relatively easy since, if $M_{\sigma_{K+1}}(0) < M_{\sigma_K}(0)$, then using the marginal payoff matrix and by induction, $M_{\sigma_{K+1}}(K-1) < M_{\sigma_K}(K-1) = \frac{c}{\beta}$. This is a contradiction to $M_{\sigma_{K+1}}(K-1) > M_{\sigma_{K+1}}(K) = \frac{c}{\beta}$. Therefore, $M_{\sigma_{K+1}}(0) \geq M_{\sigma_K}(0)$.

The marginal payoffs are bounded as follows,

$$(M_{\sigma_X}(0) + M_{\sigma_X}(X-1)) + \omega_X \sum_{k=0}^{X-1} M_{\sigma_X}(k) = b/\beta + c/\beta \quad (34)$$

However, since

$$\omega_{K+1}\sum_{k=0}^{K}M_{\sigma_{K+1}}(k) > \frac{K+1}{K}\omega_{K+1}\sum_{k=1}^{K}M_{\sigma_{K+1}}(k)$$

$$> \frac{K+1}{K}\frac{K(K+2)}{(K+1)^2}\frac{K+1}{K+2}\omega_K\sum_{k=0}^{K-1}M_{\sigma_K}(k)$$

$$= \omega_K\sum_{k=0}^{K-1}M_{\sigma_K}(k)$$

and $M_{\sigma_{K+1}}(0) + M_{\sigma_{K+1}}(K) > M_{\sigma_K}(0) + M_{\sigma_K}(K-1)$, a contradiction occurs. Therefore, for $\beta = \beta_1^L$, $M_{\sigma_{K+1}}(K) < \frac{c}{\beta}$. This means $\beta_2^L > \beta_1^L$. This completes (A).

(B) To prove $\beta_2^H > \beta_1^H$, let $\beta = \beta_1^H$, we need to show that the protocol $\Pi_{K+1}$ must have $M_{\sigma_{K+1}}(K+1) < c/\beta$. We use contradiction to prove this. The idea is: Suppose $M_{\sigma_{K+1}}(K+1) \geq c/\beta$, then we show that all the marginal payoffs of $\Pi_{K+1}$ are large enough such that they violate the restriction imposed by the bounded benefit $b$ and cost $c$.

Suppose $M_{\sigma_{K+1}}(K+1) \geq M_{\sigma_K}(K) = c/\beta$. According to the matrix equation, similar to part (A), by induction we can get,

$$\frac{M_{\sigma_{K+1}}(K+1-k)}{M_{\sigma_K}(K-k)} > \left(\frac{\omega_{K+1}+1}{\omega_K+1}\right)^k > \frac{(K+1)^3}{K(K+2)^2}, \forall 0 \leq k \leq K$$

Also $M_{\sigma_{K+1}}(0) \geq M_{\sigma_K}(0)$. The marginal payoffs are bounded as follows,

$$(M_{\sigma_X}(0) + M_{\sigma_X}(X)) + \omega_X\sum_{k=0}^{X}M_{\sigma_X}(k) = b/\beta + c/\beta \qquad (35)$$

However, since

$$\omega_{K+1}\sum_{k=0}^{K+1}M_{\sigma_{K+1}}(k) > \frac{K+2}{K+1}\omega_{K+1}\sum_{k=1}^{K+1}M_{\sigma_{K+1}}(k)$$

$$> \frac{K+2}{K+1}\frac{K(K+2)}{(K+1)^2}\frac{(K+1)^3}{K(K+2)^2}\omega_K\sum_{k=0}^{K}M_{\sigma_K}(k)$$

$$= \omega_K\sum_{k=0}^{K}M_{\sigma_K}(k)$$

and $M_{\sigma_{K+1}}(0) + M_{\sigma_{K+1}}(K+1) > M_{\sigma_K}(0) + M_{\sigma_K}(K)$, a contradiction occurs. Therefore, for $\beta = \beta_1^H$, $M_{\sigma_{K+1}}(K+1) < \frac{c}{\beta}$. This means $\beta_2^H > \beta_1^H$. This completes part (B).

(C) To prove $\beta_2^L < \beta_1^H$, wreite $\beta = \beta_1^H$. We show that $M_{\sigma_{K+1}}(K) > M_{\sigma_K}(K) = \frac{c}{\beta}$. If not, then as in (A) we must have $M_{\sigma_{K+1}}(K) \leq M_{\sigma_K}(K) = \frac{c}{\beta}$; in that case we show $M_{\sigma_{K+1}}(k) \leq M_{\sigma_K}(k)$ for $0 \leq k \leq K$. This will again violate the restrictions imposed by $b$ and $c$.

We extend the marginal payoff matrix in (33) from $K \times K$ to $(K+1) \times (K+1)$ and incorporate $M_{\sigma_K}(K)$. If $M_{\sigma_K}(K) = \frac{c}{\beta}$, such extension does not change the solution of the marginal payoffs $M_{\sigma_K}(k), \forall k \in [0, K]$. Note the new coefficient matrix has the same size of the coefficient matrix for $\sigma_{K+1}$. Suppose $M_{\sigma_{K+1}}(K) < M_{\sigma_K}(K) = \frac{c}{\beta}$. According to the matrix equation,

$$\frac{M_{\sigma_{K+1}}(K-1)}{M_{\sigma_K}(K-1)} = \frac{(\omega_{K+1} + 2)M_{\sigma_{K+1}}(K) - c/\beta}{(\omega_K + 2)M_{\sigma_K}(K) - c/\beta} < 1$$

Moreover, for $0 \leq k \leq K$ we have

$$\frac{M_{\sigma_{K+1}}(K-k)}{M_{\sigma_K}(K-k)} = \frac{(\omega_{K+1} + 1)[M_{\sigma_{K+1}}(K) + M_{\sigma_{K+1}}(K-k+1)] - c/\beta}{(\omega_K + 1)[M_{\sigma_K}(K) + M_{\sigma_K}(K-k+1)] - c/\beta}$$

By induction, $M_{\sigma_{K+1}}(k) < M_{\sigma_K}(k)$ $0 \leq k \leq K$. However, since

$$(M_{\sigma_X}(0) + M_{\sigma_X}(X-1)) + \omega_X \sum_{k=0}^{X-1} M_{\sigma_X}(k) = b/\beta + c/\beta \qquad (36)$$

Again, the left-hand side is bigger when $X = K$ than when $X = K+1$, which is a contradiction. This completes part (C).

Combining (A), (B) and (C) establishes the desired string of inequalities. The remaining conclusions of (i) follow immediately.

The argument for (ii) is very similar and left to the reader. ∎

**Proof of Theorem 8** It is convenient to first solve the following simple maximization problem:

$$\begin{aligned} \underset{0 \leq x_1, x_2 \leq 1}{\text{maximize}} \quad & E^*(x_1, x_2) = 1 - x_1 - x_2 + x_1 x_2 \\ \text{subject to} \quad & x_1(1-x_1)^K = x_2(1-x_2)^K \end{aligned} \qquad (37)$$

To solve this problem, set $f(x) = x(1-x)^K$. A straightforward calculus exercise shows that if $0 \leq x_1 \leq \frac{1}{K+1} \leq x_2 \leq 1$ and $f(x_1) = f(x_2)$ then:

(a) $x_1 + x_2 \geq \frac{2}{K+1}$, with equality achieved at $x_1 = x_2 = \frac{1}{K+1}$.

(b) $x_1 x_2 \leq \frac{1}{K+1}$, with equality achieved at $x_1 = x_2 = \frac{1}{K+1}$.

Putting (a) and (b) together shows that the optimal solution to the maximization problem (37) is to have $x_1 = x_2 = \frac{1}{K+1}$ and $\max E^* = \left(1 - \frac{1}{K+1}\right)^2$.

Now fix a protocol $\Pi = (\alpha, \sigma_K)$ and let $\eta^\Pi$ be the corresponding invariant distribution. If we take $x_1 = \mu^\Pi, x_2 = \nu^\Pi$ then our characterization of the invariant distribution shows that $f(x_1) = f(x_2)$. By definition, $\text{Eff}(\Pi) = E^*(x_1, x_2)$ so

$$\text{Eff}(\Pi) \leq \max E^* = \left(1 - \frac{1}{K+1}\right)^2$$

On the other hand, if $\alpha = K/2$ then the invariant distribution has $\eta^\Pi(k) = \frac{1}{K+1}$ for all $k$ and

$$\text{Eff}(K/2, \sigma_K) = \left(1 - \frac{1}{K+1}\right)^2 = [K/(K+1)]^2$$

Taken together, part (ii) and (iii) are proved..

Next fix a protocol $(\alpha, \sigma_K)$. Let $\lceil \alpha \rceil$ be the least integer greater than or equal to $\alpha$ and set $K^* = 2\lceil \alpha \rceil$. There are two cases to consider.

In the first case, $K \leq K^*$.

$$\text{Eff}(\alpha, \sigma_K) \leq \left(1 - \frac{1}{K+1}\right)^2 \leq \left(1 - \frac{1}{K^*+1}\right)^2 = \left(1 - \frac{1}{2\lceil \alpha \rceil + 1}\right)^2$$

which is the desired result in the first case.

In the second case, $K > K^*$. Define the protocol $\Pi' = (\lceil \alpha \rceil, \sigma_K)$; let $\eta'$ be the invariant token distribution for $\Pi'$. Let $\Pi^* = (\lceil \alpha \rceil, \sigma_{K^*})$; note that the invariant token distribution $\eta^*$ is uniform ($\eta^*(k) = \frac{1}{K^*+1} = \frac{1}{2\lceil \alpha \rceil + 1}$ for all $k = 0, 1, ..., K^*$). Note that $\Pi'$ and $\Pi$ have the same strategy component but that the token supply for $\Pi'$ is larger than for $\Pi$, and that $\Pi'$ and $\Pi^*$ have the same token supply but that the strategy component of $\Pi'$ has a higher threshold.

We assert that $\eta'(0) \geq \frac{1}{2\lceil\alpha\rceil+1}$. If not then $\eta'(0) < \frac{1}{2\lceil\alpha\rceil+1} = \frac{1}{K^*+1}$. It follows that for all $k \in \{0, 1, ..., K\}$ we have $\eta'(k) < \frac{1}{K^*+1} = \eta^*(k)$. Hence

$$\lceil\alpha_0\rceil = \sum_{k=0}^{K^*} k\eta^*(k) = \sum_{k=0}^{K^*} k(\eta^*(k) - \eta'(k)) + \sum_{k=0}^{K^*} k\eta'(k)$$

$$\leq K^* \sum_{k=0}^{K^*} (\eta^*(k) - \eta'(k)) + \sum_{k=0}^{K^*} k\eta'(k) = K^*(1 - \sum_{k=0}^{K^*} \eta'(k)) + \sum_{k=0}^{K^*} k\eta'(k)$$

$$= K^* \sum_{k=K^*}^{K} \eta'(k) + \sum_{k=0}^{K^*} k\eta'(k) \leq \sum_{k=K^*}^{K} k\eta'(k) + \sum_{k=0}^{K^*} k\eta'(k) = \lceil\alpha_0\rceil$$

This is a contradiction. Hence, $\eta'(0) \geq \frac{1}{2\lceil\alpha\rceil+1}$.

Because the token supply for $\Pi$ is less than $\Pi'$, the number of agents with no tokens is larger, so $\eta(0) > \eta'(0) \geq \frac{1}{2\lceil\alpha\rceil+1}$. Hence

$$\text{Eff}(\Pi) = (1 - \eta(0))(1 - \eta(K)) < (1 - \eta(0)) < \left(1 - \frac{1}{2\lceil\alpha\rceil+1}\right)$$

which is the desired result in the second case. This complete the proof for part (i).

∎

**Proof of Theorem 9** Both assertions follow immediately by combining Theorems 7 and 8. ∎

**Proof of Theorem 10** We first derive the lower bound $K^L$. If $\Pi_K = (K/2, \sigma_K)$ is an equilibrium protocol then consecutive marginal utilities bear the relationship

$$\phi_l M_{\sigma_K}(k-1) + \phi_c M_{\sigma_K}(k) = -\phi_r M_{\sigma_K}(k+1) > 0$$

(Because $\beta$ is fixed, we suppress it in the notation.) Therefore, $M_{\sigma_K}(k) > \frac{-\phi_l}{\phi_c} M_{\sigma_K}(k-1)$. By induction,

$$M_{\sigma_K}(k) > \left(\frac{-\phi_l}{\phi_c}\right)^k M_{\sigma_K}(0) > \left(\frac{\rho\beta}{2(1-\beta) + 2\rho\beta}\right)^k M_{\sigma_K}(0)$$

Because $\phi_c M_{\sigma_K}(0) = (1-\nu)\rho b - \phi_r M_{\sigma_K}(1) > (1-\nu)\rho b + (1-\nu)\rho c$, we have

$$M_{\sigma_K}(0) > \frac{(1-\nu)\rho b}{\phi_c} = \frac{\rho\beta}{2(1-\beta)+2\rho\beta}\frac{b+c}{\beta}$$

Therefore,

$$M_{\sigma_K}(k) > \left(\frac{\rho\beta}{2(1-\beta)+2\rho\beta}\right)^{k+1}\frac{b+c}{\beta} \tag{38}$$

Because $\Pi_K$ is assumed to be an equilibrium protocol, we must have $M_{\sigma_K}(K) \leq c/\beta$. Moreover, we must also have

$$\left(\frac{\rho\beta}{2(1-\beta)+2\rho\beta}\right)^{K+1}\frac{b+c}{\beta} \leq \frac{c}{\beta}$$

because otherwise $M_{\sigma_K}(K) > c/\beta$. Therefore,

$$K \geq \max\{\log_{\frac{\rho\beta}{2(1-\beta)+2\rho\beta}}\frac{c}{b+c} - 1, 0\} \tag{39}$$

This provides the lower bound $K^L$.

We now derive the upper bound $K^H$. Rewriting the relation between consecutive marginal utilities we obtain

$$0 = \phi_l M_{\sigma_K}(k-1) + \phi_c M_{\sigma_K}(k) + \phi_r M_{\sigma_K}(k+1)$$
$$> \phi_l M_{\sigma_K}(k-1) + (\phi_c + \phi_r)M_{\sigma_K}(k)$$

Therefore, $M_{\sigma_K}(k) < \frac{-\phi_l}{\phi_c+\phi_r}M_{\sigma_K}(k-1)$. By induction,

$$M_{\sigma_K}(k) < \left(\frac{-\phi_l}{\phi_c+\phi_r}\right)^k M_{\sigma_K}(0) < \left(\frac{\rho\beta}{1-\beta+\rho\beta}\right)^k M_{\sigma_K}(0)$$

Because $\phi_c M_{\sigma_K}(0) = (1-\nu)\rho b - \phi_r M_{\sigma_K}(1) < (1-\nu)\rho b - \phi_r b/\beta = 2(1-\nu)\rho b$, we have,

$$M_{\sigma_K}(0) < \frac{\rho\beta}{1-\beta+\rho\beta}\frac{2b}{\beta}$$

Therefore,

$$M_{\sigma_K}(k) < \left(\frac{\rho\beta}{1-\beta+\rho\beta}\right)^{k+1}\frac{2b}{\beta} \tag{40}$$

Because $\Pi_K$ is assumed to be an equilibrium protocol, we must have $M_{\sigma_K}(K-1) \geq c/\beta$. Moreover,

$$\left(\frac{\rho\beta}{1-\beta+\rho\beta}\right)^K \frac{2b}{\beta} \geq \frac{c}{\beta}$$

because otherwise $M_{\sigma_K}(K-1) < c/\beta$. Therefore,

$$K \leq \log_{\frac{\rho\beta}{1-\beta+\rho\beta}} \frac{c}{2b} \tag{41}$$

This provides the upper bound $K^H$.

Combining the two estimates yields the range containing all integers $K$ for which $\Pi_K$ is an equilibrium protocol. The estimate for efficiency follows immediately since $Eff(\Pi_K) \geq Eff(\Pi_{K^L})$ if $K \geq K^L$, so the proof is complete. ∎